\documentclass[conference,compsoc]{IEEEtran}
\usepackage{graphicx}
\usepackage[dvipsnames]{xcolor}
\usepackage{multirow}
\usepackage{algorithm}
\ifCLASSOPTIONcompsoc
  \usepackage[nocompress]{cite}
\else
  \usepackage{cite}
\fi
\ifCLASSINFOpdf
\else
\fi
\usepackage{amsmath}

\usepackage{algorithmic}
\ifCLASSOPTIONcompsoc
 \usepackage[caption=false,font=footnotesize,labelfont=sf,textfont=sf]{subfig}
\else
 \usepackage[caption=false,font=footnotesize]{subfig}
\fi
\usepackage{url}
\usepackage{tikz}
\usepackage{eso-pic}

\begin{document}
\AddToShipoutPictureFG*{%
\begin{tikzpicture}[remember picture, overlay]

% Accepted statement above title.
% Adjust the yshift value if needed.
\node[
    anchor=north,
    font=\normalsize,
    align=center
] at ([yshift=-1.1in]current page.north)
{Accepted to IEEE Symposium on Security and Privacy 2026};

% Copyright box at bottom of first page.
\node[
    anchor=south,
    draw,
    line width=0.4pt,
    inner sep=4pt,
    text width=0.92\textwidth,
    align=left,
    font=\footnotesize
] at ([yshift=0.42in]current page.south)
{\textcopyright2026 IEEE.  Personal use of this material is permitted.  Permission from IEEE must be obtained for all other uses, in any current or future media, including reprinting/republishing this material for advertising or promotional purposes, creating new collective works, for resale or redistribution to servers or lists, or reuse of any copyrighted component of this work in other works.};

\end{tikzpicture}%
}
\title{EyeSpy: Inferring Eye Gaze via Side-Channel Attacks Against Foveated Rendering}

\author{
\IEEEauthorblockN{
Paul Maynard, Harris Amjad, Camila Molinares, Bo Ji, and Brendan David-John
}
\IEEEauthorblockA{
Department of Computer Science\\
Virginia Tech\\
Blacksburg, VA, USA\\
\{pmaynard, harris98, camilavm, boji, bmdj\}@vt.edu
}
}


% compsoc conferences compared to traditional conferences), use this


\maketitle

%need to look for small changes that were not highlighted as well--- like spelling fixes that were made.
\begin{abstract}
While eye tracking provides valuable capabilities for virtual reality, such as gaze interaction and dynamic foveated rendering (DFR), eye-tracking data can inadvertently reveal sensitive user information if not properly protected. Current protections, such as adding permission prompts or gatekeeping gaze data, are insufficient on DFR-enabled systems because gaze data is used internally to drive DFR. When DFR is implemented, objects in the fovea (i.e., immediate gaze area) incur a higher GPU workload than those in the periphery. This gaze-contingent workload creates a novel side channel, which can be leveraged to reconstruct gaze positions. Specifically, we design a novel attack that sweeps imperceptible high-cost objects (HCOs) across the user's field of view and logs rendering performance metrics (e.g., frame rate or frame time) commonly exposed through standard game engines. Then, we correlate variation in these metrics (caused by HCO-foveal overlap) with the known HCOs' positions to infer gaze coordinates directly without using eye-tracking APIs. Our experimental results show that mean gaze prediction errors (1.1--4.4~degrees) across the Meta Quest Pro, Varjo XR-4, and desktop platforms are comparable to typical eye-tracker accuracy. We demonstrate that the attack generalizes across various hardware platforms, standard game engines, and foveated rendering pipelines. Finally, we design defense mechanisms based on supervised and unsupervised detectors that can flag the attack reliably (F1 of 0.99) over short time windows. 
\end{abstract}

\IEEEpeerreviewmaketitle

\section{Introduction} \label{Introduction}
Virtual reality (VR) headsets are increasingly being used in a wide array of domains such as entertainment~\cite{meta_entertainment}, education~\cite{lampropoulos2024virtual}, training~\cite{mergen2024reviewing}, and military~\cite{zyda2005visual}. Many modern headsets now include eye tracking to enable features such as gaze-interface control~\cite{pfeuffer2017gaze+}, user intent modeling~\cite{narkar2024gazeintent, david2021towards}, and dynamic foveated rendering (DFR)~\cite{nvidia_vrs}.

Eye tracking provides valuable capabilities for VR but can inadvertently leak sensitive user information if not protected~\cite{kroger2020does}. Prior work shows that eye movements reveal medical conditions, cognitive and emotional states, and personality traits~\cite{koch2023you,bozkir2023eye,kroger2020does}. They have also been exploited for manipulation and behavioral targeting~\cite{Ramirez2024dec,Rucco2024from}, used to profile users for advertising~\cite{krauss2024makes}, and linked to greater phishing susceptibility~\cite{lin2019susceptibility}. While many of these risks rely on high-rate gaze streams, even low-rate gaze can support inference of cognitive load and personal preferences~\cite{kroger2020does,mahanama2022eye}. Generally, low-rate gaze data reveals which objects and regions receive sustained user attention over time; consequently, it remains sensitive even when coarse. Therefore, ensuring the security of eye-tracking data is critical.

Modern VR headsets expose eye tracking to applications at different levels of detail: some provide continuous raw gaze streams, while others restrict apps to coarse, heavily processed data, or disable eye-tracking Application Programming Interfaces (APIs) entirely. Prior work has proposed defenses that regulate these channels using permission prompts, gatekeeper-style APIs, or injected noise~\cite{david2021privacy,steil2019privaceye,li2021kalvarepsilonido,liu2019differential}. However, all of these mechanisms treat gaze as an explicit data channel that an app requests \emph{directly}.

We revisit this eye-tracking access landscape and show that even when eye-tracking APIs are restricted or disabled, gaze can still leak through rendering behavior \emph{indirectly}.
Specifically, we show that as long as a system (i) uses gaze internally for DFR and (ii) exposes rendering performance metrics (e.g., frame rate or frame time) to apps, a malicious app can infer gaze by exploiting how DFR affects GPU workload and, in turn, app-accessible performance metrics. We show that a novel side-channel attack can circumvent existing protections by exploiting DFR-induced variation in rendering to infer privacy-relevant gaze information.

\begin{figure}[!t]
\centering
\includegraphics[width=1.0\linewidth]{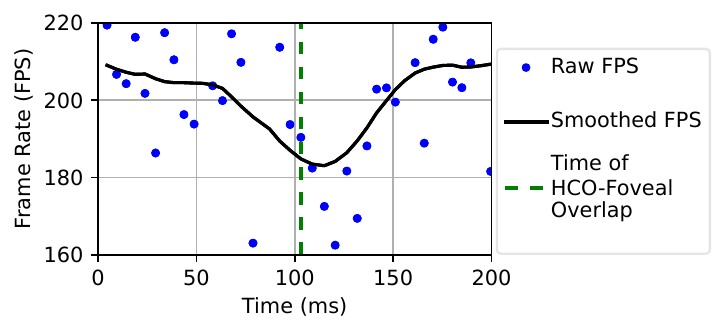}
\caption{Frame rate drops as a high-cost object (HCO) moves across the foveal region. The drop tracks increasing HCO-foveal overlap; the dashed green line marks the time of maximum overlap.}
\label{fig:HCO_Foveal}
\end{figure}

We target DFR, a common GPU optimization technique that renders a user's foveal region at a higher shading rate and the periphery at a lower rate~\cite{patney2016towards,swafford2016user,meng2018kernel} (see details in Sec.~\ref{FoveatedRendering}). These spatially varying shading rates create a \emph{correlation}: rendering a given object incurs more GPU work when it falls in the fovea than in the periphery. When an object moves relative to gaze, this location-dependent workload produces measurable variation in app-accessible performance metrics (e.g., frame rate or frame time). By controlling high-cost object (HCO) movement, we leverage this effect to infer gaze. As illustrated in Fig.~\ref{fig:HCO_Foveal}, introducing a fully transparent HCO that sweeps across the user’s field of view (FOV) creates synchronized variation in these metrics when the HCO overlaps the fovea. By logging an affected rendering performance metric and correlating its variation with the HCO’s position over time, we infer time-stamped gaze coordinates without accessing raw eye-tracking data.

In VR apps, such leakage could reveal which avatars, prompts, advertisements, or other scene elements repeatedly receive a user's attention, even without explicit interaction. For example, a social VR lobby app could present politically, medically, or personally sensitive cues and use inferred gaze fixations (or avoidance patterns) to profile user interests or mental state, enabling targeted advertising or phishing even without eye-tracking permissions. One example tied to user interests is financial harm: if cues related to debt relief or short-term loans elicit disproportionately longer fixations or a higher fixation frequency compared to other stimuli, the app could infer that the user has financial concerns and then target the user with advertisements for financial services. A second example, tied to mental state, is a health-related harm: if dieting or rapid-weight-loss cues elicit similar aberrant fixation behavior, the app could infer heightened body-image concern or distress and use that signal to target the user with personalized advertisements. 

However, realizing this novel attack and achieving high-accuracy gaze inference poses three interrelated challenges:

\textbf{(C1) Designing imperceptible but high-impact scan HCOs.}
The scan object must add enough GPU work in the foveal region to produce a measurable change in app-accessible performance metrics, yet keep the experience smooth and the attack imperceptible to the user (i.e., no visible artifacts, judder, or input lag). If the object is too costly to render, frame-rate dips and stutter expose the attack; if it is not costly enough, the resulting variation is buried in background noise and inference accuracy collapses. HCO scan time introduces a three-way tradeoff between (i) how frequently the attacker can obtain gaze estimates, (ii) how many metric samples are available per scan to average out noise and provide finer HCO positions per frame, and (iii) how far the user’s gaze can move while the HCO traverses the FOV. Shorter scan times yield more frequent estimates than longer scan times but fewer frames per scan, and thus unstable position estimates; longer scan times provide more samples and denser coverage, but less frequent gaze estimates and increased chance that gaze shifts during the scan and blurs the inferred location. Selecting scan geometry and timing is therefore a complex part of HCO design and directly affects the feasibility of the next two challenges.

\textbf{(C2) Extracting workload variation under dynamic rendering conditions.}
Even with a tuned HCO, per-frame performance metrics are noisy and time-varying due to background load, head motion, animation, and asynchronous VR scheduling. The attacker must recover a weak latency-shifted workload pattern caused by HCO-foveal overlap (Fig.~\ref{fig:HCO_Foveal}), suppress confounds and false positives from unrelated scene changes, and convert that pattern into an accurate estimate of the HCO angle at the time of maximum overlap. In short, while \textbf{(C1)} governs strength of induced workload variation, \textbf{(C2)} focuses on isolating that variation and mapping it to time-stamped gaze coordinates.

\textbf{(C3) Generalizing the attack across hardware platforms, game engines, and foveation pipelines.}
Portability further complicates extracting workload variation. Across game engines and hardware platforms, both the behavior of the available metrics and the point at which they are sampled in the rendering pipeline differ. Gaze-contingent foveation may be managed internally, such as with hardware Variable Rate Shading (VRS), or implemented at the application level through shading-rate maps or eye-tracked view configurations. These platform- and runtime-level configurations affect (i) whether increased HCO-foveal overlap produces a dip or a peak in the metric, (ii) the latency of the metric response relative to the overlap, (iii) the effective sampling rate and stability of the logged metrics, and (iv) the coordinate mapping from screen space to scan angle. Consequently, the attacker needs an inference pipeline that is robust to such platform-specific differences and can be retuned with lightweight calibration, rather than reengineered from scratch for each engine-platform configuration.

To that end, we introduce a practical attack that reconstructs gaze data from DFR-induced workload variation without eye-tracking APIs or privileged access, demonstrate cross-platform generalization, and provide lightweight detectors. Our main contributions are three-fold:

\textbf{Foveated-rendering side-channel attack.}
We identify and exploit a new gaze-leakage side channel unique to DFR: transparent HCOs scanned across the FOV increase GPU workload when they pass through the foveal region, exposing gaze position through app-accessible performance metrics. We design a radar-inspired, scan-based attack that uses imperceptible HCOs, sequential horizontal and vertical scans with controlled scan times, simple smoothing, extremum (a local maximum or minimum, depending on performance metric) detection on smoothed data, and calibrated offsets to convert metric traces into gaze estimates without eye-tracking APIs or privileged access. This combination of HCO design, scan scheduling, and metric processing yields a practical, unprivileged attack that addresses \textbf{(C1)} on HCO design and scan scheduling as well as \textbf{(C2)} on extracting workload variation from noisy metrics in VR.

\textbf{Cross-platform evaluation in VR and desktop settings.}
We evaluate the same attack on three representative platforms: two head-mounted displays, Meta Quest Pro (MQP) (stand-alone; Unity) and Varjo XR-4 (PC-tethered; Unreal), and a desktop setting (Godot). The mean angular errors (X and Y) are $4.36^\circ$ and $2.88^\circ$ on MQP, $3.98^\circ$ and $3.24^\circ$ on XR-4, and $1.23^\circ$ and $0.96^\circ$ on the desktop. These results are comparable to typical VR eye-tracker accuracy (2--3$^\circ$)~\cite{aziz2024evaluation}, which demonstrates generalization across game engines and foveation pipelines, addressing \textbf{(C3)}.

\textbf{Lightweight attack detection.}  
We propose and evaluate simple supervised (logistic regression) and unsupervised (K-means) detectors that identify scanning behavior from short windows of performance metric statistics. On desktop, logistic regression trained on labeled data achieves an F1 of 0.93, while a label-free K-means model reaches an F1 of 0.99 once the detection window meets or exceeds the scan time. The same configuration generalizes well to VR. The detectors use features distinct from the attack pipeline (e.g., standard deviation and an outlier proportion rather than extrema timing), making runtime screening practical. We also discuss how adaptive scan strategies could reduce detectability, and directions for hardening such defenses.
\section{Background}
\label{Background}
We briefly review three foundations for this work: privacy risks of eye-tracking data, how DFR optimizes display quality and performance, and how rendering performance metrics can inadvertently leak information via side channels.
\subsection{Privacy Risks of Eye-Tracking Data}
\label{Eye Tracking & Eye-Gaze Data}
Eye-tracking data enable many useful features, but can also expose sensitive information~\cite{david2021privacy,garrido2023sok}. The growing deployment of eye tracking in VR has led the community to scrutinize the privacy implications of collecting and using gaze data in immersive environments~\cite{kroger2020does,selinger2023eye,bozkir2023eye}.
 
Gaze data comprise fixations (low velocity) and rapid movements (e.g., saccades, smooth pursuits, and microsaccades)~\cite{lappi2016eye}. Both categories of eye movements can be used for authentication~\cite{lohr2022eye}, re-identification~\cite{george2016score,david2021privacy}, and to reveal personal and demographic attributes about the user~\cite{kroger2020does, koch2023you}. Fixation-based features, in particular, support a wide range of inferences (see Table~\ref{tab:fixation_inference}) and can be reliably extracted from gaze data even at low sampling rates~\cite{drews2024strategies}. Longer fixations are typically interpreted as reflecting deeper cognitive processing of the fixated content, so identifying when and where they occur can reveal which regions of a scene receive sustained visual attention~\cite{mahanama2022eye}. While all studies in Table~\ref{tab:fixation_inference} analyze privacy risks of gaze data from non-immersive settings, these risks generalize to eye movements broadly and remain relevant to gaze data in VR, since fixation patterns in both settings arise from the same fundamental processes of visual attention and cognition.

Recent work further shows that eye movements and other behavioral signals collected in VR can enable user identification and profiling. 
For example, BehaVR demonstrates that combinations of VR sensor streams, including eye gaze, can uniquely identify users based on behavioral interaction patterns~\cite{jarin2025behavr}. 
Similarly, Aziz \textit{et al.} show that gaze and motion data can enable user re-identification in VR and that privacy protections applied to only a subset of sensor streams may still leave users exposed through other available signals~\cite{aziz2025exploring}. 
These findings suggest that recovering gaze information, even indirectly through side channels, can materially increase privacy risks in immersive environments.

\begin{table}[!t]
    \centering
        \caption{Inferences from fixation data in prior work.}
    \begin{tabular}{|l|}
        \hline
        {\centering \textbf{Information Inferred from Fixations}} \\
        \hline
        Cognitive load and mental workload~\cite{tao2020eye, guo2021eye, chen2013automatic, vcenvek2020cultural, hayhoe2004advances} \\
        \hline
        Personality traits (e.g., Big Five traits)~\cite{hoppe2018eye, chen2022eye} \\
        \hline
        Emotional responses and affective states~\cite{tao2020eye, chua2005cultural, capriola2021attention, usee2020abstract} \\
        \hline
        Neurological conditions~\cite{tao2020eye, moghadami2021investigation, li2024automating} \\
        \hline
        Preferences and interests~\cite{orquin2013attention, wedel2008eye, chen2023effect} \\
        \hline
        Cultural affiliations~\cite{vcenvek2020cultural, chua2005cultural}
        \\ \hline       
    \end{tabular}
    \newline
    \label{tab:fixation_inference}
\end{table}

Prior work has proposed privacy mechanisms that either regulate access to eye-tracking data~\cite{david2021privacy} or add noise to protect the gaze stream~\cite{steil2019privaceye, li2021kalvarepsilonido, liu2019differential}. These mechanisms assume that gaze is requested \emph{directly} from the eye tracker via an API and focus on controlling or perturbing that specific stream. They do not address the possibility that gaze could be inferred \emph{indirectly} from other app-accessible signals such as performance metrics affected by gaze-contingent rendering load. Our work empirically demonstrates that such a side channel exists under typical DFR pipelines, indicating that protecting gaze-data access alone is insufficient to mitigate gaze-based privacy risks.

\subsection{Foveated Rendering}
\label{FoveatedRendering}
Human foveal vision spans about $6^\circ$ around one's gaze position~\cite{heun2012perifoveal}. Visual acuity drops steeply with eccentricity as cone density falls off in the periphery~\cite{barris2005vision}. Thus, only a small region of a display can be viewed in high detail at one time, while display resolution can be much lower in the periphery without any perceived differences~\cite{patney2016towards}.

Foveated rendering leverages this limitation of the human vision system to optimize GPU performance by rendering only the foveal region in high resolution, while the periphery is rendered at a lower resolution. This reduces rendering time while preserving perceived quality, which is critical for VR headsets that must sustain high frame rates to avoid simulator sickness~\cite{wang2023effect, conner2022virtual, zhang2023s}. Wang \textit{et al.} show that 90~FPS is an important threshold to meet or exceed to reduce simulator sickness in headsets~\cite{wang2023effect}, and utilizing DFR makes it easier to achieve this target frame rate.

\begin{figure}[!t]
\centering
\includegraphics[width=1.0\linewidth]{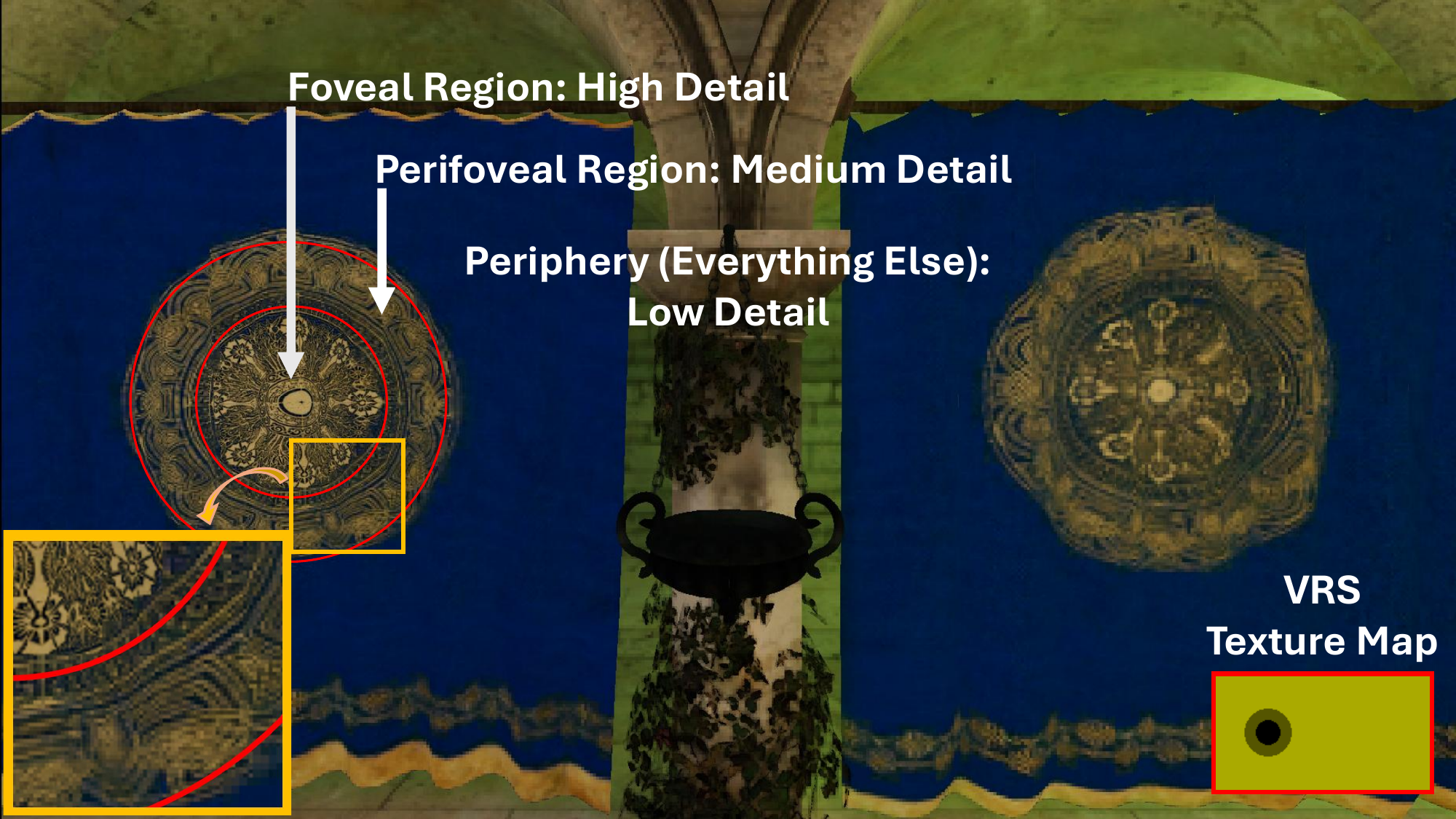}
\caption{An example of a VRS texture map used for DFR. The foveal and accompanying perifoveal regions dynamically shift based on eye-tracking data, optimizing GPU resources to render the user's gaze area with the highest level of detail.
}
\label{fig:VRSOverlay}
\end{figure}

Foveated rendering is typically implemented using VRS where the GPU varies the resolution of pixel groups across the display~\cite{nvidia_vrs}. The resolution or shading rate is specified using a texture map and allows the GPU to assign one color to an entire group of pixels. Fig.~\ref{fig:VRSOverlay} illustrates high, medium, and low detail regions corresponding to the foveal, perifoveal, and peripheral regions specified in the texture map. DFR maintains a high-detail foveal region that follows the user's gaze by frequently updating this texture map with eye-tracking data~\cite{albert2017latency}. Since shading rate directly affects GPU workload, rendering an object in the foveal region rather than the periphery incurs greater GPU cost. As a result, certain performance metrics (e.g., frame rate or frame time) reflect a higher rendering load when content appears near the user’s gaze.

\subsection{GPU Side-Channel Attacks}
Game and VR app developers use performance metrics to profile and optimize system performance. Profiler APIs provide access to hardware metrics and pipeline statistics such as memory allocation, vertex count, thread usage, and frame rate to ensure experiences run efficiently across rendering systems~\cite{guo2011practical,zhang2023s}. Although intended for debugging and optimization, these metrics in VR can reflect user behavior and leak behavioral signals to third-party apps, making them a natural target for side-channel attacks~\cite{zhang2023s}.

Prior work has demonstrated a wide range of GPU side-channel exploitation in desktop settings, such as identifying running apps~\cite{naghibijouybari2018rendered}, estimating user inputs~\cite{wang2019cracking, zhang2023s}, inferring website activity~\cite{mai2023novel}, and even recovering encryption keys~\cite{naghibijouybari2018rendered}. In immersive settings such as VR, GPU metrics have been used to infer gesture input and detect nearby bystanders~\cite{zhang2023s}. To our knowledge, no research has explored side-channel attacks targeting DFR to reconstruct time-stamped gaze coordinates via performance metrics.
\section{Threat Model} \label{ThreatModel} 
This section defines a threat model that describes the attacker’s objective, capabilities, and system assumptions. We assume that the attacker is motivated to obtain gaze information from an arbitrary VR user to perform re-identification or profiling. Profiling information puts the user at risk for manipulative personalization~\cite{mhaidli2021identifying,Ramirez2024dec} and could support targeted advertisements or otherwise financially benefit the attacker~\cite{adams2018ethics,mhaidli2023shockvertising,mhaidli2025intriguing}.
\subsection{Attacker's Objective and Capabilities}
A user discovers the attacker's third-party VR app in a commercial app store and installs it. The app passes review because it behaves like an ordinary game, and the attack logic uses only standard engine features (e.g., transparent meshes and frame-level profiling APIs). In some ecosystems, such as Meta App Lab, the attacker could also distribute the app through channels with even lighter review requirements than the commercial app stores~\cite{meta_appstore}.

\begin{figure}[!t]
\centering
\includegraphics[width=1.0\linewidth]{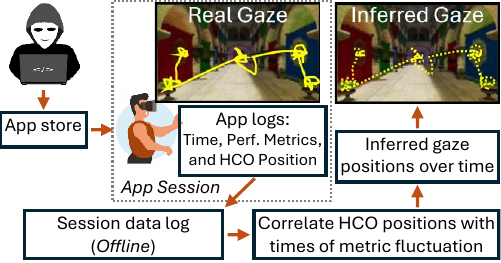}
\caption{Threat model: An attacker-controlled app logs performance metrics and HCO positions at each frame, then correlates metric fluctuations with HCO positions to infer the user’s time-stamped gaze coordinates, without direct access to eye-tracking data.}
\label{fig:ThreatModel}
\end{figure}

After starting the app session, the app behaves as advertised in the store (e.g., a VR game), while code in the app periodically renders HCOs in the active scene and samples a performance metric (e.g., frame rate or frame time) exposed by the game engine. The app logs the metric value and the HCO’s position each frame. After the session, the attacker correlates metric fluctuations with HCO positions to reconstruct time-stamped gaze coordinates (see Fig.~\ref{fig:ThreatModel}).

\textbf{Goal.} The attacker aims to infer a time series of the user’s gaze coordinates during ordinary play without using eye-tracking APIs or privileged access.

\textbf{Capabilities.} The attacker controls all assets and logic in the app: the app can place, animate, and set the complexity of the HCOs; the app can also read per-frame performance metrics commonly available via game engine APIs intended for performance tuning~\cite{zhang2023s}. The attacker needs only standard 3D development skills with common game engines (e.g., Unity, Unreal, or Godot) and XR SDKs (e.g., OpenXR).

\textbf{Scope.} All scanning and measurements occur within the attacker’s foreground app. The app does not overlay or inspect other apps and does not access system-level eye-tracking streams. In our experiments, we log ground-truth gaze via the headset SDK solely for calibrating spatial offsets before app deployment, and for offline evaluation; the attack pipeline itself operates only on the performance metrics and known HCO positions. We do not implement downstream profiling inferences on the reconstructed gaze; instead, we refer to prior work showing that fixation-level gaze features and upsampled gaze trajectories can support such inferences~\cite{kroger2020does,koch2023you,bozkir2023eye,jiao2023supreyes,mahanama2022eye}.

\subsection{System Assumptions and Attack Surface}
We assume the target device is a VR headset that supports DFR as in modern systems such as the MQP, Apple Vision Pro (AVP), and Varjo XR-4. Real-time gaze coordinates are used internally by the headset to direct GPU resources so the fovea is rendered at higher detail using VRS~\cite{patney2016towards} or other DFR approaches~\cite{meng2018kernel,meng2020eye, openxr_quadviews}.

Table~\ref{tab:HeadsetLandscape} summarizes representative headsets and the range of eye-tracking access models exposed to apps. MQP and Varjo XR-4 provide relatively raw gaze streams (gated by permission prompts or system configuration) while still using gaze internally for DFR. AVP, in contrast, does not expose continuous gaze to apps; instead, it shares only event-level eye-based interaction (e.g., which element is selected) while using raw gaze data internally for interaction and DFR~\cite{apple_eyes_hands_privacy}. Galaxy XR adopts an intermediate, gatekeeper style in which apps request coarse or fine eye-tracking permissions, with DFR requiring fine access~\cite{unity_foveated_rendering}.

\begin{table}[t]
\centering
\caption{App-accessible eye-tracking and performance-metric access on representative VR headsets.}
\setlength{\tabcolsep}{4pt}\renewcommand{\arraystretch}{1.1}
\begin{tabular}{|l|c|c|c|c|}
\hline
\textbf{Headset} &
\begin{tabular}[c]{@{}c@{}}App consent\\ for DFR\end{tabular} &
\begin{tabular}[c]{@{}c@{}}Gaze\\ API\end{tabular} &
\begin{tabular}[c]{@{}c@{}}Data\\ granularity\end{tabular} &
\begin{tabular}[c]{@{}c@{}}Perf. metric\\ collection\end{tabular} \\
\hline
Meta Quest Pro   & Yes   & Yes & Raw & Allowed \\
Varjo XR-4  & No    & Yes & Raw & Allowed \\
Apple Vision Pro  & No    & No  & N/A & Allowed \\
Samsung Galaxy XR   & Yes    & Yes & Coarse/Fine\footnotemark{} & Allowed\\
\hline
\end{tabular}
\label{tab:HeadsetLandscape}
\end{table}
\footnotetext{``Fine'' is the term Samsung uses for precise gaze comparable to ``Raw'' on other headsets.}
In our threat model, we focus on the stricter case in which raw eye-tracking streams are restricted or not exposed to third-party apps, while the same gaze signal still drives DFR and standard per-frame performance metrics remain accessible~\cite{apple_eyes_hands_privacy}. Among current devices, AVP exemplifies this restricted case. We use MQP and Varjo XR-4 as experimental platforms representative of standalone and tethered systems that support DFR and permit ground-truth gaze logging for offline evaluation. However, the attack relies only on performance metrics in a deployment scenario.

The attack surface consists of two elements: gaze-contingent foveation and app-accessible per-frame performance metrics. Thus, the attack is feasible whenever DFR is enabled, and the malicious app can observe performance metrics affected by DFR. As HCOs periodically sweep across the user’s FOV and pass through the foveal region, they induce detectable GPU load variation in certain performance metrics. The attacker treats DFR as a black box: they do not observe gaze or shading rates directly, only the effect of gaze-contingent shading on certain performance metrics.

Finally, we assume that scan times are on the order of hundreds of milliseconds. In our experiments, we use scan windows of up to 500~ms because, at typical VR frame rates (60--200~FPS), this provides enough samples per scan to localize the workload extremum while keeping the scan short relative to typical visual fixation durations. Note that 500~ms is not a strict requirement for the attack: the usable scan time depends on factors such as frame rate and performance metric noise. We therefore treat 500~ms as a practical operating point rather than a minimum or maximum bound. This temporal resolution is sufficient to demonstrate that a malicious app can reconstruct gaze coordinates from performance metrics. We discuss the privacy implications of such inferred gaze data in Sec.~\ref{Discussion}. This threat model reflects realistic developer capabilities and a low barrier to distribution, making the attack a credible threat.
\section{System Design} \label{System Design}
This section describes how the attack induces controlled GPU-workload variation via gaze-contingent foveation and converts app-accessible performance metrics into gaze estimates. We first present the key idea and motivation, then the system architecture and workflow, followed by HCO design and scan logic, and conclude with the inference pipeline.

\subsection{Key Insight and Design Motivation}
Our system induces controlled variation in GPU workload via DFR and exploits the resulting app-accessible performance metric side channel (e.g., frame rate or frame time) to infer gaze without calling eye-tracking APIs. When an HCO overlaps the foveal region, rendering cost increases and produces a localized variation in the side-channel metric. The attack exploits this behavior by synchronizing HCO movement with metric logging and mapping the time of the smoothed metric extremum to the HCO’s logged position, yielding an estimate of the gaze coordinate without direct access to eye-tracking data.

We initially explored a static grid approach in which multiple HCOs were rendered simultaneously at fixed screen locations, each with distinct rendering cost. The idea was to correlate observed GPU load with the region containing the user's gaze. However, this method proved ineffective: to achieve meaningful spatial resolution, the number of HCOs required significantly reduced overall frame rate. Additionally, simultaneously rendering many HCOs made it difficult to isolate individual contributions to GPU load, reducing signal clarity and effectiveness of gaze inference.

Instead, we adopt a spatio-temporal design inspired by radar tracking systems~\cite{faa2024aim}. Rather than a static grid, we sweep a single HCO horizontally or vertically across the FOV. The sweep yields a localized extremum in the metric when the HCO crosses the fovea; we later map the extremum’s timestamp to the HCO’s position at that time to obtain one axis of gaze.

Achieving high-accuracy gaze inference in this way faces two key design challenges. First, HCOs must produce detectable performance variation while remaining imperceptible and avoiding any degradation in user experience (\textbf{C1}). Second, the system must extract consistent signal patterns from noisy performance metrics despite the variability introduced by rendering pipelines, head motion, and latency in VR environments \textbf{(C2)}. The following subsections describe how we address these challenges in our design.
\subsection{System Overview}
Our attack is implemented as an ordinary VR app that runs in the foreground on the target device and does not call eye-tracking APIs. Once launched, the app renders invisible HCOs inside its own scene and samples per-frame performance metrics exposed by the game engine. 

The app executes a repeating sequence of \emph{scans}. For each scan, a single HCO sweeps across the FOV along one axis while the app logs its screen-space position and the per-frame performance metric. We refer to the duration of one sweep as the \emph{scan time} ($T_{\text{scan}}$).

\begin{figure}[!t] 
\centerline{\includegraphics[width=1.0\linewidth]{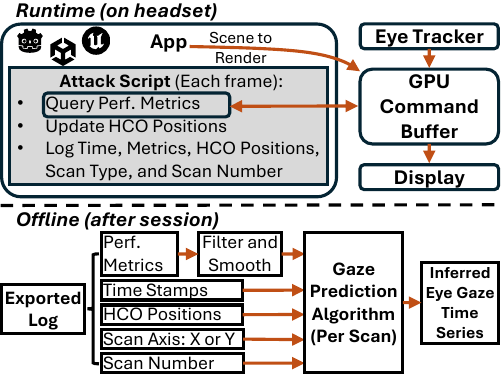}}
\caption{Attack Design: During runtime (top), the app logs per-frame performance metrics, HCO position, timestamp, scan axis (X or Y), and scan number. Scan numbers group frames into scan windows corresponding to a single HCO sweep. Offline (bottom), the metric extremum within each window is identified, and its timestamp is mapped to the HCO position at that moment to infer gaze over time.}
\label{fig:DesignModell}
\end{figure}

Fig.~\ref{fig:DesignModell} summarizes the architecture. The top panel shows the runtime behavior inside the app: a script schedules scans, moves the HCOs, and logs the per-frame metric and HCO positions while the user plays. The bottom panel shows the offline stage, where the exported log is passed through the inference pipeline to produce a gaze time series. Algorithm~1 summarizes the per-scan inference pipeline. The same architecture can run on desktop or in VR; only the attacked metric and offset calculations differ by hardware platform.

In the following sections, we describe the key design components that enable this system: how we construct and control HCOs to create clean inference signals (Sec.~\ref{Design of HCOs}), and how we process GPU performance metric data to infer gaze position from timing patterns (Sec.~\ref{Infer Gaze Positions}).

\subsection{HCO Design and Scan Logic}
\label{Design of HCOs}

At runtime, the app moves HCOs along pre-configured trajectories across the user's FOV to induce metric variation during HCO-foveal overlap. This section focuses on how these HCOs are constructed and controlled to reliably induce measurable variation in the performance metric without affecting the user’s visual experience, addressing \textbf{(C1)}.

Each HCO is constructed from many overlapping transparent cylinders arranged to span the full height (for horizontal scans) or full width (for vertical scans) of the display; this ensures that each HCO intersects gaze along its scanned axis regardless of where the user looks. The number of cylinders is selected through offline tuning to generate sufficient workload for a measurable signal without degrading overall performance.
 HCOs are rendered fully transparent to remain visually imperceptible to the user.

Our scanning process is inspired by air traffic control (ATC) radar systems. ATC radars rotate while emitting radio waves and receive reflections when aircraft are in the direction of emission~\cite{faa2024aim}. When the radar receives a reflection from an aircraft, the direction of the aircraft can be inferred from the radar's azimuth at that time. Analogously, our system scans HCOs across the user's FOV in a timed, repeating trajectory. When an HCO intersects the user's foveal region, it induces a spike in GPU workload, much like a radar reflection, allowing us to infer gaze direction from the timing of the performance fluctuation. 

\begin{figure}[!t] 
\centerline{\includegraphics[width=1.0\linewidth]{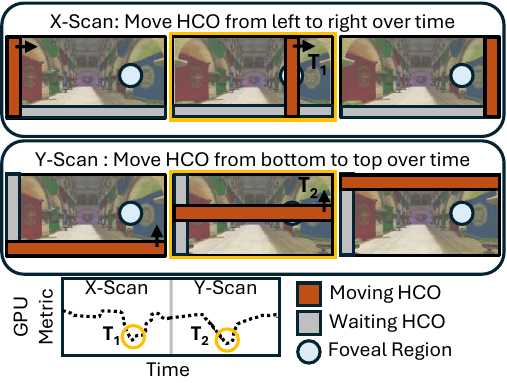}}
\caption{Attack process: One HCO scans while the other waits. Dips occur at $T_1$ and $T_2$ during HCO-foveal overlap for the X- and Y-scans respectively. The HCOs are invisible but colored here for illustration.}
\label{fig:AttackProcess}
\end{figure}

The attack script alternates X- and Y-axis scans in a loop: X-scans move one HCO left to right; Y-scans move the other HCO bottom to top as in Fig.~\ref{fig:AttackProcess}. Each frame, the app logs the HCO’s position and the chosen GPU metric, creating a precise mapping from HCO position to metric response. When the HCO crosses the fovea, the metric exhibits a localized extremum. The inference module uses these within-scan extrema to recover gaze coordinates.

\subsection{Gaze Inference and Metric Processing}
\label{Infer Gaze Positions}
The app logs an entry each frame with the performance metric value and HCO position, and we process the logged data offline after the session. Each scan spans many frames and is analyzed independently from other scans. 

Formally, let each frame $k$ in the session log record
a timestamp $t_k$, a performance metric value $m_k$ (e.g., frame
time or frame rate), the HCO angular position $\theta_k$ along the
active scan axis, an axis label $a_k \in \{\mathrm{X},\mathrm{Y}\}$,
and a scan identifier $\text{scan}_k$. The offline pipeline groups
frames by scan identifier and, for each scan window, extracts a single
gaze estimate along that axis by (i) smoothing the metric trace,
(ii) locating the within-scan extremum corresponding to HCO-foveal
overlap, and (iii) mapping the extremum time to the HCO position and
applying a spatial correction. Algorithm~1
summarizes this procedure.

\begin{algorithm}[t!]
\caption{Scan-Based Gaze Inference} 
\label{alg:scan-inference}
\begin{algorithmic}[1]
\STATE \textbf{Input:} Session log
       $\{(t_k, m_k, \theta_k, a_k, \text{scan}_k)\}_{k=1}^K$,
       where $t_k$ is the frame timestamp, $m_k$ the performance metric,
       $\theta_k$ the HCO angular position, $a_k \in \{\mathrm{X},\mathrm{Y}\}$
       the scan axis, and $\text{scan}_k$ a scan identifier.
\STATE \textbf{Output:} Time-stamped gaze samples  $\{(t^{(j)}, x^{(j)}, y^{(j)})\}_j$.
\STATE Group frames into scan windows
       $\mathcal{W} = \{w_1, \dots, w_{N_\text{scan}}\}$ using
       $\text{scan}_k$ and axis labels.
\STATE Initialize an empty list of per-scan estimates $\mathcal{S}$.
\FOR{each scan window $w \in \mathcal{W}$ (in chronological order)}
  \STATE Let $\{t_{w,j}, m_{w,j}, \theta_{w,j}\}_{j=1}^{n_w}$ be the
         timestamps, metric values, and HCO positions in scan $w$,
         and let $a_w \in \{\mathrm{X},\mathrm{Y}\}$ denote its scan axis.
  \STATE Smooth the remaining $\{m_{w,j}\}$
         to obtain $\tilde m_{w,j}$.
  \STATE Determine the extremum index
         \[
           j_w^\star =
           \begin{cases}
             \arg\min_j \tilde m_{w,j}, & \text{if higher load decreases } m,\\[2pt]
             \arg\max_j \tilde m_{w,j}, & \text{if higher load increases } m.
           \end{cases}
         \]
  \STATE Initial gaze estimate along axis $a_w$:
         $P_i(w) \gets \theta_{w,j_w^\star}$ at time
         $t_i(w) \gets t_{w,j_w^\star}$.
  \STATE Apply the appropriate spatial correction for axis $a_w$
         to obtain a final position $P_f(w)$ from $P_i(w)$.
  \STATE Append $(t_i(w), P_f(w), a_w)$ to $\mathcal{S}$.
\ENDFOR 

// At this point, $\mathcal{S} = \{(t_s, P_f(s), a_s)\}_{s=1}^{N_\text{scan}}$ is ordered in time, with scans alternating X then Y.

\STATE Initialize an empty list of gaze samples $\mathcal{G}$.
\FOR{$j = 1$ \TO $N_\text{scan} / 2$}
  \STATE Let $s \gets 2j - 1$ \COMMENT{X-scan}, $s' \gets 2j$ \COMMENT{Y-scan}.
  \STATE Let $(t_s, P_f(s), a_s)$ and $(t_{s'}, P_f(s'), a_{s'})$ be
         the $s$-th and $(s')$-th elements of $\mathcal{S}$.
  \STATE Set $t^{(j)} \gets \tfrac{1}{2}\bigl(t_s + t_{s'}\bigr)$.
  \STATE Set $x^{(j)} \gets P_f(s)$ and $y^{(j)} \gets P_f(s')$.
  \STATE Append $(t^{(j)}, x^{(j)}, y^{(j)})$ to $\mathcal{G}$.
\ENDFOR
\STATE \textbf{return} $\mathcal{G}$.
\end{algorithmic}
\end{algorithm}

Because rendering cost propagates to the metric with a small delay, we apply a spatial offset correction per axis to better align the estimate with true gaze. In VR, this correction is eccentricity-dependent; on desktop, it is a constant offset (see Sec.~\ref{sec:vr_user_study} and Sec.~\ref{sec:desktop_results}). Pairing consecutive X- and Y-scan estimates yields one $(x,y)$ sample per scan cycle, reconstructing a time-stamped gaze series across the session.

This approach directly addresses \textbf{(C2)}. It isolates a weak, latency-shifted signal from noisy per-frame metrics using lightweight filtering and extrema selection. Rather than complex runtime inference, we use statistical processing offline. The attack operates solely on readily available system metrics, such as frame rate or render time, which are commonly exposed by standard game engines. While prior work~\cite{zhang2023s} has demonstrated similar metric-based side channels in VR, none have targeted gaze or exploited DFR in this way.
\section{Implementation} 
\label{Implementation}
In this section, we describe how we implement our scan-based side-channel attack on a stand-alone VR headset (MQP, using Unity), a desktop-tethered VR headset (Varjo XR-4, using Unreal), and a desktop setting (Godot). We focus on three aspects: how DFR is implemented, how HCOs are constructed and controlled, and how per-frame performance metrics are processed to infer gaze. Prior work shows that a variety of app-accessible metrics (e.g., object count, memory usage, or frame time) can be used as side channels~\cite{zhang2023s}, but here we focus on simple per-frame timing metrics because, in preliminary exploration, they produced the clearest and most consistent response to HCO-induced workload changes across platforms. On MQP we log frame time via Unity’s \texttt{Time.deltaTime}; on Varjo XR-4 we use Unreal’s GPU frame time; and on desktop we derive frame rate from Godot’s timing callbacks.

\subsection{VR Implementation}
We implement the attack on a stand-alone MQP VR headset, which serves as our primary platform. We also implemented the attack on a desktop-tethered Varjo XR-4 using Unreal (utilizes desktop GPU for headset display), but we highlight the implementation of the more popular consumer-facing MQP. Details for Varjo XR-4 implementation can be found in Appendix~\ref{app:VR}. The VR pipeline follows the scan-based design described in Sec.~\ref{System Design}: HCOs sweep across the FOV while the app logs per-frame performance metrics and HCO position. The next three subsections detail the MQP environment, the foveation source, the attack mechanics, and inference procedure.

\subsubsection{MQP VR Environment}
\begin{figure}[t]
  \centering
  \includegraphics[width=\columnwidth]{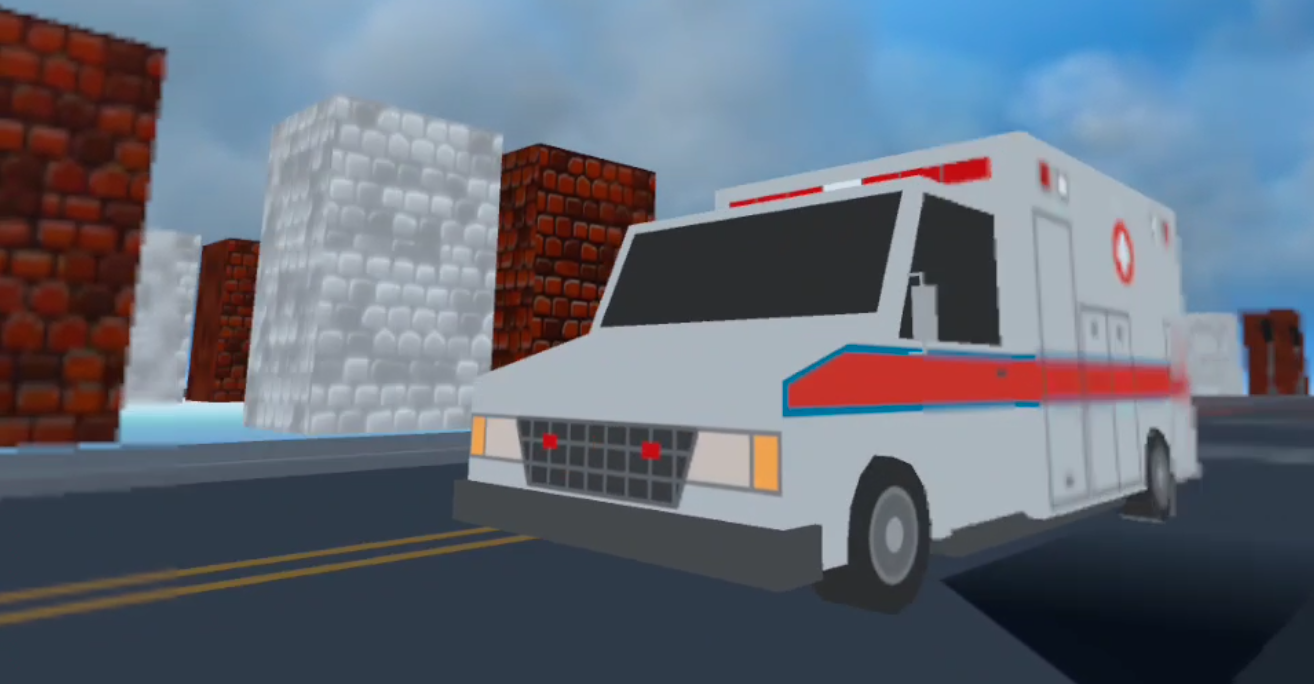}
  \caption{MQP “city street” environment used in the user study. Proxy “buildings” (brick-textured cubes) line the roadway; a low-poly vehicle moves at a fixed speed to provide a controlled dynamic element.}
  \label{fig:mqp_scene}
\end{figure}

We implemented the VR build in Unity~6 (using Universal Render Pipeline) for stand-alone execution on MQP via OpenXR. The user study was implemented in a test scene consisting of a minimal ``city street'': a straight roadway with sidewalks and traffic lights, brick-textured building proxies (static cubes) spaced along the road, and a vehicle that moves up and down the street as a controlled dynamic element (Fig.~\ref{fig:mqp_scene}). This configuration provides object density, occlusion, and motion typical of interactive VR. We use this scene as a controlled test environment to assess gaze inference feasibility under realistic VR rendering conditions. This study scene is not itself the motivating privacy-sensitive scenario discussed earlier; it serves as a simplified streamlined environment for evaluating whether the attack can infer gaze in practice.

We render stereo views from a single perspective camera (FOV $\approx 108^\circ$ horizontal and $96^\circ$ vertical). HCOs are rendered at a fixed, close distance from the camera and their thickness scaled with viewing angle as they move across the FOV so that their screen-space thickness remains constant approximately $10^\circ$, reducing perspective-induced variability in HCO-foveal interaction. Each frame, we log the game's running time, scan number, Unity’s accessible \texttt{Time.deltaTime} (frame time), and the scanning HCO’s angular position (azimuth/elevation, in degrees) relative to the camera’s forward vector. All instrumentation uses app-accessible engine features and does not require privileged profilers or developer shell access.

\subsubsection{VR Foveated Rendering}
On MQP we rely on platform foveation exposed via OpenXR. When eye-tracked foveation is used, the foveal region follows gaze. The attack requires only that a foveal region be rendered at higher quality and cost than the periphery. When the HCO overlaps the fovea, the additional shading work increases frame time (equivalently, decreases frame rate) relative to when the HCO is in the periphery. This mechanism is vendor-agnostic and works similarly on all headsets leveraging DFR.

\subsubsection{VR Attack and Gaze Inference}
We executed the MQP attack with tuned parameters: 16 cylinders per HCO and $T_{\text{scan}} = 500$~ms per axis. Each participant completed two 60~s runs: one with the attack active, and one with the attack disabled (HCOs locked to the FOV margins without moving). Participants were not told about the attack. After each run, they completed a short questionnaire to assess whether they noticed any visual artifacts or responsiveness differences; we report those results in Sec.~\ref{Results}.

\begin{figure}[t]
\centering
\subfloat[Raw performance metric]{%
  \includegraphics[width=0.49\linewidth]{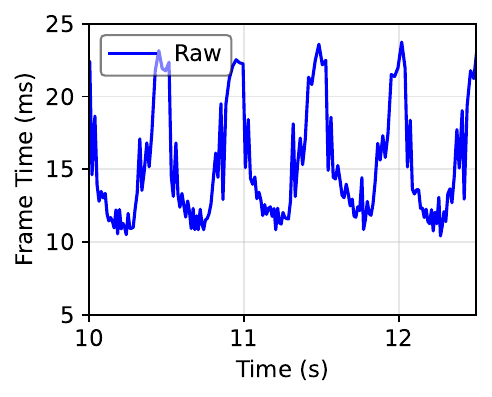}%
  \label{fig:subfigMPQ7RAW}}
\hfill
  \subfloat[Smoothed performance metric]{%
  \includegraphics[width=0.49\linewidth]{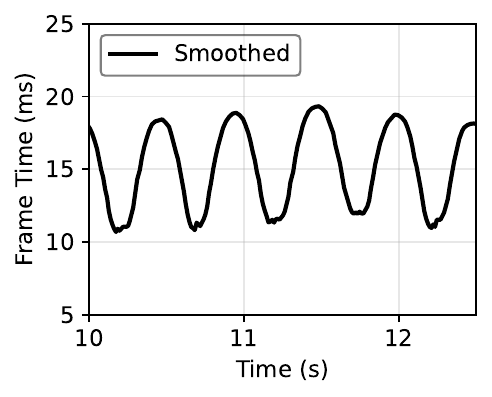}%
  \label{fig:subfigMPQ7SMOOTH}}
\hfill
\subfloat[X-estimate (without offset)]{%
  \includegraphics[width=0.49\linewidth]{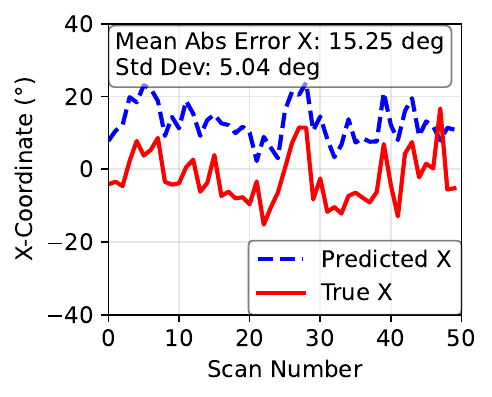}%
  \label{fig:subfigXNoOff}}
\hfill
\subfloat[X-prediction (with offset)]{%
  \includegraphics[width=0.49\linewidth]{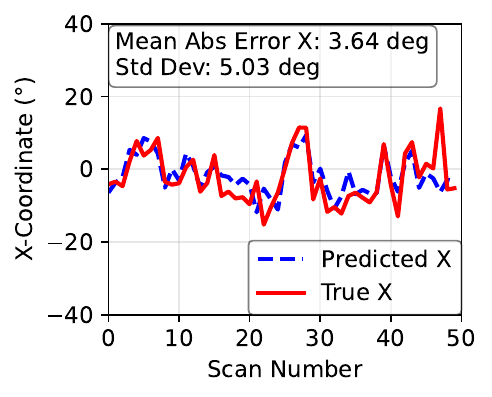}%
  \label{fig:subfigXOff}}
\hfill
  \subfloat[Y-estimate (without offset)]{%
  \includegraphics[width=0.49\linewidth]{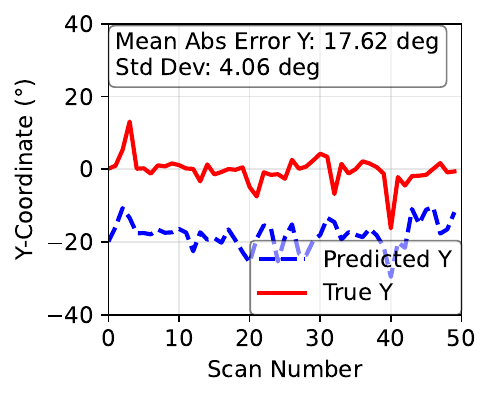}%
  \label{fig:subfigYNoOff}}
\hfill
\subfloat[Y-prediction (with offset)]{%
  \includegraphics[width=0.49\linewidth]{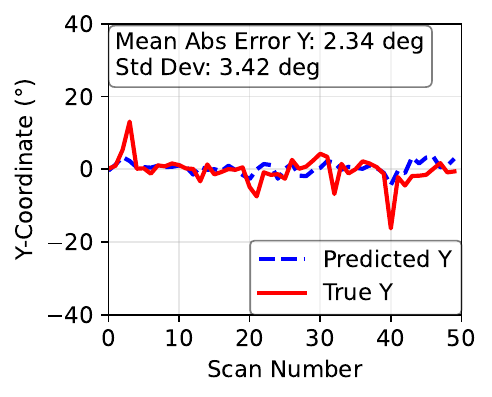}%
  \label{fig:subfigYOff}}
\caption{Data-processing progression on MQP (Desktop is the same). Raw frame-time data (a) is smoothed via a SG filter to produce a clean peak per scan (b). Selecting the maximum smoothed frame time per scan and using the HCO position reported at that time yields an initial gaze estimate (c, e). Applying the learned offset produces the final prediction (d, f).}
\label{fig:vrPrelimResultsFree_MQP}
\end{figure}

On MQP we implement the steps in Algorithm~1 with frame time as the performance metric, processing each scan window independently. Fig.~\ref{fig:vrPrelimResultsFree_MQP} illustrates the process:

\noindent \textbf{Smooth.} We smooth the raw data using an SG filter~\cite{savitzky1964smoothing} (window size 27, polynomial order 2).

\noindent \textbf{Estimate gaze.} We locate the within-scan extremum in the smoothed series (maximum for frame time), then map its timestamp to the HCO’s logged position to obtain an initial gaze estimate coordinate along that axis.

\noindent \textbf{Apply spatial offset.} Finally, we apply an eccentricity-based correction to the estimate to obtain the final prediction.

We process scans sequentially. Pairing the X- and Y-axis estimates from consecutive scans yields one $(x,y)$ per scan cycle and reconstructs a time-stamped gaze trajectory. For evaluation, we compare against the best per-axis ground truth: for the X-scan, we use the horizontal component of true gaze when the HCO is nearest foveal center; for the Y-scan, we use the analogous vertical component. This does not produce a single instant where both axes are measured simultaneously, but it gives the most precise per-axis ground truth available under the sequential scan design.

\subsection{Desktop Implementation}
In addition to the VR implementation, we instantiate the same attack pipeline in a desktop setting using Godot. This allows us to explore generalization to a different device, engine, and DFR approach, and to study how different frame rates affect attack accuracy under controlled conditions.

\subsubsection{Desktop Environment}
We ran experiments on an Alienware Aurora R15 with an Intel Core i9-13900KF, 64~GB RAM, and an NVIDIA RTX~3090 GPU. To reproduce frame-rate ranges typical of VR scenes with moderate complexity, we limited GPU power to 250~W, yielding stable targets at 120, 160, and 200~FPS, depending on scene complexity. Although this setting weakened the attack signal, it was necessary to reduce the desktop frame rate to realistic VR-representative ranges. All desktop experiments used Godot~4.2 on Ubuntu 22.04~LTS.

We implemented our desktop experiments in Godot because it is common to both traditional and VR game development. We used the Sponza scene, a standard rendering benchmark with complex lighting, textures, and meshes~\cite{godot-sponza}. The scene camera spanned a FOV of $107.52^\circ$ horizontally, and $75^\circ$ vertically.

\subsubsection{Desktop Foveated Rendering}
We emulate DFR on desktop without an eye tracker by driving the GPU’s VRS with a precomputed sequence of ``shading-rate maps.'' Each map is a low-resolution image whose pixel values set the shading rate across the screen: 1$\times$1 in the fovea, 2$\times$2 in the perifovea, 4$\times$4 in the periphery (Fig.~\ref{fig:VRSOverlay} shows a sample). To mimic gaze-contingent foveation, we take a gaze trace and, at 100~Hz, generate the corresponding map that centers the high-detail foveal region on that gaze point~\cite{adhanom2023eye}. We use a 192$\times$108 map with a 1920$\times$1080 framebuffer: the map controls shading rate across the whole screen (it does not reduce output resolution). The foveal disk is 21 pixels in diameter (about 11.8$^\circ$ horizontally, 14.6$^\circ$ vertically); the perifoveal ring spans an additional 20 pixels; the remainder is periphery. We advance the map every 10~ms (100~Hz), with each rendered frame using the most recent map.

Each experiment selects one gaze sequence and loads the precomputed VRS maps with the app. We evaluate gaze traces from two real VR datasets (DGaze and ET-DK2).

\noindent\textbf{DGaze.} The DGaze dataset~\cite{hu20_dgaze} records human gaze during VR navigation and visual-search tasks. Participants freely explored visually rich 3D scenes, producing natural sequences of fixations and saccades. The released gaze coordinates are already normalized and smoothed; we therefore use a representative subject’s gaze trace as provided and do not impose additional constraints on eye-movement patterns.

\noindent\textbf{ET-DK2.} The ET-DK2 dataset~\cite{brendan_david_john_2021_4642612} contains raw eye tracking during free viewing of $360^\circ$ images in VR. Participants were instructed to explore each panoramic scene naturally, again yielding unconstrained sequences of fixations and saccades under varying visual content. We use 18 of the 19 subjects (Subject~7 data was unavailable) to assess generalization across subjects and across diverse gaze trajectories.

\subsubsection{Desktop Attack and Gaze Inference}
\label{sec:desktop_attack} We executed the attack with tuned parameters (see Sec.~\ref{sec:desktop_results}): 250~W GPU power cap, 300~cylinders per HCO, and $T_{\text{scan}} = 200$~ms per axis. These settings were fixed across desktop experiments.

Each run lasted 30~s. To stabilize baseline frame rate at 120, 160, or 200~FPS, we added invisible full-FOV rectangular meshes (“backdrops”) until the scene reached the target frame rate. Backdrops uniformly increased load without masking HCO-induced variation.

All attack assets and scripts were grouped under an \texttt{AttackScene} node in the Godot scene. The main \emph{perspective} camera rendered the scene with the HCOs masked out. A dedicated \emph{orthographic} camera rendered only the HCOs, maintaining constant screen-space width as they swept. We logged per-frame frame rate, scan number, HCO screen position, and ground-truth gaze.

On desktop, we implemented Algorithm~1 with frame rate as the performance metric. Each scan window is processed independently, and the same as the VR implementation with a few minor exceptions: (i) outlier filtering was implemented, (ii) we applied an additional smoothing after the SG filter by averaging each point with its four preceding and four following neighbors, and (iii) the offset was constant but chosen based on average frame rate.

We process scans sequentially, pairing the X- and Y-axis estimates from consecutive scans to obtain one $(x,y)$ prediction per scan cycle and reconstruct a time-stamped gaze trajectory. Per-axis ground truth is obtained in the same way as in the VR implementation.
\section{Evaluations} \label{Results}
This section presents a quantitative evaluation of our attack in both VR and desktop environments, analyzing gaze prediction accuracy and generalizability across gaze behavior, frame rates, and hardware. The VR environment uses real-time headset gaze for DFR, while the desktop environment uses simulated DFR with per-frame VRS maps driven by gaze sequences. In both environments, a scan sweeps one HCO across a single axis with fixed duration $T_{\text{scan}}$; pairing consecutive X- and Y-axis scans yields one $(x,y)$ sample per scan cycle. VR inferences use frame time as the performance metric, while desktop inferences use frame rate. To compensate for the latency between HCO-foveal interaction and the recorded metric, we apply spatial offsets to initial position estimates: an eccentricity-dependent correction in VR, and a constant per-axis offset in the desktop setting derived from calibration trials. Unless noted, $T_{\text{scan}}$ is 500~ms (VR) and 200~ms (desktop).
\subsection{VR Evaluation}
\label{sec:vr_user_study}

We conducted an IRB-approved user study on the MQP to evaluate attack performance under real-time eye tracking. Fifteen participants provided informed consent. The Simulator Sickness Questionnaire (SSQ)~\cite{kennedy1993simulator} was administered post-session; no participants reported symptoms.

Each participant completed two 60~s ``free-look'' runs in which they freely explored the scene without prompts, producing natural head and eye movements. One run used the attack in the \emph{active} state and the other in the \emph{inactive} state. In the inactive state, the HCOs were rendered but fixed at the screen margins, preventing intersections with the foveal region. After each run, participants completed a brief questionnaire about visible artifacts and usability to assess any perceptible impact of the attack.

On MQP we apply the per-scan preprocessing and gaze-estimation pipeline described in Sec.~\ref{Implementation} and Algorithm~1, using frame time as the performance metric. Each scan window is processed independently to produce an initial gaze estimate along the scanned axis, and we set $T_{\text{scan}} = 500$~ms per axis to accommodate the effective frame rates on MQP.

To correct the residual latency-induced bias and eccentricity-dependent distortion, we apply an angle-dependent linear offset. For each axis we learn coefficients $(a,b)$ such that the corrected estimate is
\[
P_f = P_i - (a \,\theta + b),
\]
where $P_i$ is the initial estimate and $\theta$ is the HCO’s angular position (horizontal for X-scans, vertical for Y-scans) at the metric extremum. Coefficients are trained and evaluated in a leave-one-out cross-validation (LOOCV) scheme for each participant: in each fold, we grid-search $(a,b)$ on data from $N{-}1$ participants to minimize mean absolute error along that axis, then apply the resulting coefficients to the held-out participant. Reported errors are averaged over all folds.

\textbf{Accuracy.} Table~\ref{tab:leave_one_out_MQP} shows LOOCV accuracy on MQP and Varjo XR-4.
\begin{table}[!t]
\centering
\caption{Mean and SD of absolute gaze error for MQP and Varjo XR-4 (leave-one-participant-out evaluation)}
\begin{tabular}{|l|c|c|}
\hline
\textbf{Headset} & \textbf{X Error ($^\circ$)} & \textbf{Y Error ($^\circ$)} \\ \hline
\textbf{MQP}      & $4.36\,(\pm\,5.89)$ & $2.88\,(\pm\,3.60)$ \\ \hline
\textbf{Varjo XR-4} & $3.98\,(\pm\,4.81)$ & $3.24\,(\pm\,4.18)$ \\ \hline
\end{tabular}
\label{tab:leave_one_out_MQP}
\end{table}
Across both headsets, mean absolute errors are  approximately 4$^\circ$ in X and 3$^\circ$ in Y, with relatively large standard deviations reflecting natural head and eye motion during free viewing. Subfigures~\subref{fig:subfigXOff} and~\subref{fig:subfigYOff} of
Fig.~\ref{fig:vrPrelimResultsFree_MQP} illustrate how the inferred gaze position follows the ground truth data by plotting data from one participant with coefficients trained on all others. The magnitude of these errors and SDs indicates room for improvement, but also shows that the attack is feasible and yields estimates in the same order as typical HMD eye-tracking noise. Prior work on gaze super-resolution and scanpath reconstruction~\cite{jiao2023supreyes} suggests that such coarse reconstructions can, in principle, be refined and used as inputs to downstream analyses, including sensitive inferences and re-identification attacks~\cite{aziz2023assessing}.

\begin{figure}[!t]
\centering
\subfloat[X-axis]{%
  \includegraphics[width=0.48\linewidth]{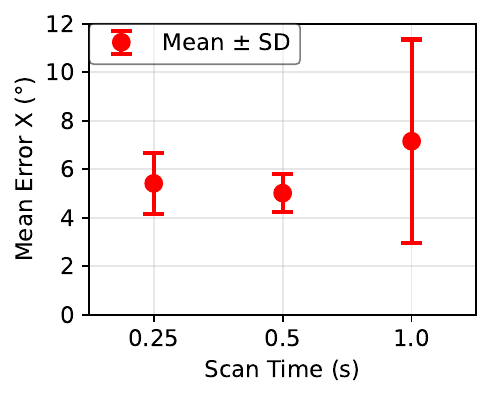}%
  \label{fig:XaxisMQP}}
\hfill
\subfloat[Y-axis]{%
  \includegraphics[width=0.48\linewidth]{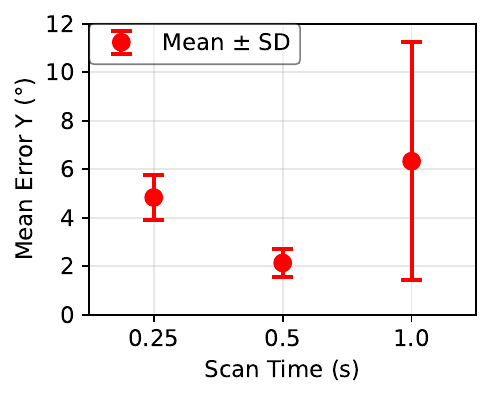}%
  \label{fig:YaxisMQP}}
\caption{Scan time vs.\ gaze-estimation error on Meta Quest Pro. Dots show mean absolute error across five participants; error bars indicate $\pm$1~SD. Each participant repeated the same scenario with scan times of 0.25~s, 0.5~s, and 1.0~s. In both X-axis (a) and Y-axis (b), 0.5~s was optimal.}
\label{fig:ScanTimeErrorVR}
\end{figure}

We also ran the same setup with 5 of the MQP participants, but with 0.25~s and 1.0~s scan times. These runs also went through a LOOCV evaluation and confirmed that the 0.5~s scan time was optimal (at least out of 0.25, 0.5, and 1.0~s options) as we concluded in pilot runs (see Fig.~\ref{fig:ScanTimeErrorVR}).

\textbf{Sensitivity to MQP foveation strength.}
To characterize within-system sensitivity to DFR configuration, we repeated the MQP evaluation for a new participant while varying the runtime foveation-strength setting in Unity. For each setting, the participant completed five full sessions, each consisting of eye calibration, 60 seconds of city-street exploration, and headset removal before the next session. The average offset values from the prior participants were used, and no other parameters were adjusted except for the foveation strength. Table~\ref{tab:MQPFoveationStrength} shows that gaze error increases as foveation strength is reduced from the default value of 1.0 to 0.5 and 0.1. The results are expected: weaker foveation reduces the shading-rate difference between the foveal and peripheral regions and therefore weakens the workload signal induced by HCO-foveal overlap, resulting in higher errors. Nonetheless, the side channel remains observable and continues to support coarse gaze inference under reduced-strength settings.

\begin{table}[!t]
\centering
\caption{Mean absolute gaze error on MQP at different foveation-strength settings for one participant.}
\begin{tabular}{|r|c|c|}
\hline
\textbf{Foveation Strength} & \textbf{X Error ($^\circ$)} & \textbf{Y Error ($^\circ$)} \\
\hline
\textbf{(default) 1.0} & $2.83\,(\pm\,0.62)$ & $2.72\,(\pm\,0.56)$ \\ \hline
\textbf{0.5} & $5.35\,(\pm\,0.94)$ & $3.63\,(\pm\,2.05)$ \\ \hline
\textbf{0.1} & $9.27\,(\pm\,1.25)$ & $7.60\,(\pm\,3.14)$ \\ \hline
\end{tabular}
\label{tab:MQPFoveationStrength}
\end{table}

\textbf{Perceptibility.} After each run, participants answered four 0--100 slider questions about visual artifacts, visual quality, and the presence of strange objects. We compared responses between the \emph{attack-active} and \emph{attack-disabled} runs for the same participant. On average, participants reported low levels of visual artifacts in both conditions (16.2 vs. 20.6), and rated the overall visual quality similarly (52.7 vs. 54.9). When asked whether they noticed any strange objects that did not match the rest of the VR environment, responses remained near the bottom of the scale (4.6 vs. 0.1). We did not observe any statistically significant differences between the attack and baseline runs on these items. Taken together, these results suggest that our HCO design does not introduce obvious visual artifacts or degrade perceived visual quality, addressing \textbf{(C1)}.

\subsection{Desktop Evaluation}
\label{sec:desktop_results}
We tuned attack parameters in the Sponza scene using central-fixation trials as calibration to set scan time, HCO dimensions, and a per-axis spatial offset. The selected parameters for all desktop trials were: GPU power cap 250~W to reproduce ideal VR frame-rate ranges on desktop, $T_{\text{scan}}=200$~ms, HCO width $5.3^\circ$ (X) and height $6.9^\circ$ (Y), and 300 co-located cylinders per HCO.

Using the same central-fixation trials, we computed constant spatial offsets for each axis and frame-rate target to correct systematic prediction error. Offsets were derived by averaging the difference between initial gaze estimates and the ground-truth gaze position across all scans. Table~\ref{tab:FPS_Offsets} in Appendix~\ref{app:scan-offset} summarizes the resulting offsets for each frame-rate target (120, 160, and 200~FPS), which were applied across all gaze conditions in our evaluation.

\begin{figure}[!t]
\centering
\subfloat[Without offset]{%
  \includegraphics[width=0.48\linewidth]{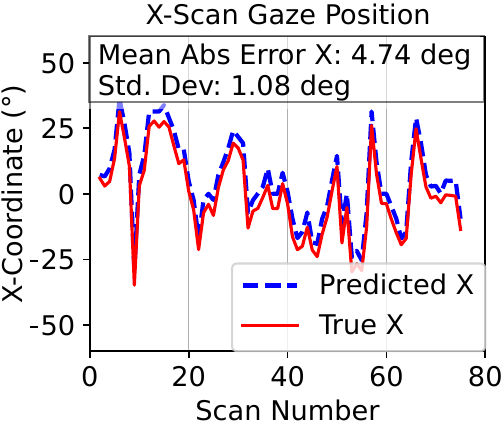}%
  \label{fig:subfigNoOff}}
\hfill
\subfloat[With offset]{%
  \includegraphics[width=0.48\linewidth]{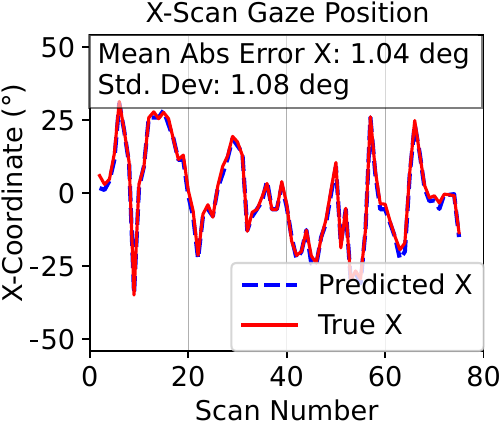}%
  \label{fig:subfigOff}}
\caption{Predicted and true X-coordinates for each scan in the ET-DK2 condition at a 200~FPS baseline. 
We report mean absolute error and standard deviation averaged over scans. 
(a) Initial X-component estimate; (b) final X-component after applying the offsets in Table~\ref{tab:FPS_Offsets}.}
\label{fig:XCoordPredictions}
\end{figure}
We evaluated both DGaze and ET-DK2 gaze-trace conditions at three frame-rate targets, with five trials per target. Gaze error is the absolute angular distance between the predicted and ground-truth gaze positions per scan; offsets from Table~\ref{tab:FPS_Offsets} in Appendix~\ref{app:scan-offset} are applied. Fig.~\ref{fig:XCoordPredictions} illustrates the improvement for one ET-DK2 trial, where raw estimates had a mean error of $\approx 4.7^\circ$ and applying the offset reduced it to $\approx 1.0^\circ$. This improvement was typical: across conditions with the same baseline frame rate, the same offset values generalized well, producing accurate gaze predictions.

\textbf{Accuracy. }
Table~\ref{tab:ErrorByFPSSummary} summarizes errors across all experiments. Mean X/Y errors were under $3.5^\circ$/$1.6^\circ$ at 120~FPS and decreased with higher frame rates, reaching $1.19^\circ$/$0.91^\circ$ at 200~FPS. These results demonstrate that accurate gaze inference is feasible using only app-accessible performance metrics, and average errors were within the 2--3$^\circ$ range of typical eye-tracking noise~\cite{aziz2024evaluation, schuetz2022eye}. Detailed error summaries for each gaze condition and frame-rate target, including subject-level results for ET-DK2, are provided in Tables~\ref{tab:ErrorWithOffset120FPS}, \ref{tab:ErrorWithOffset160FPS}, and \ref{tab:ErrorWithOffset200FPS} in Appendix~\ref{app:scan-offset}.

\begin{table}[!t]
\centering
\caption{Mean and SD of absolute gaze error in the desktop evaluation at 120, 160, and 200~FPS.
}
\begin{tabular}{|r|c|c|}
    \hline
    \textbf{FPS} & \textbf{X Error ($^\circ$)} & \textbf{Y Error ($^\circ$)} \\
    \hline
    120	&3.50 ($\pm 6.51$)	&1.59 ($\pm 2.26$) \\
    \hline
    160	&1.65 ($\pm 2.90$)	&1.12 ($\pm 1.75$) \\
    \hline
    200	&1.19 ($\pm 1.66$)	&0.91 ($\pm 1.47$) \\
    \hline
    
\end{tabular}
%}
\label{tab:ErrorByFPSSummary}
\end{table}
\begin{table}[!t]
\centering
\caption{Mean and SD of absolute gaze error in the desktop evaluation at 120~FPS (varying foveal diameter).}
\begin{tabular}{|r|c|c|}
    \hline
    \textbf{Diameter (FOV-X$^\circ$ / FOV-Y$^\circ$)} & \textbf{X Error ($^\circ$)} & \textbf{Y Error ($^\circ$)} \\
    \hline
    \textbf{5.6 / 6.9}	&2.76 ($\pm 0.89$)	&2.67 ($\pm 0.78$) \\
    \hline
    \textbf{(Default) 11.8 / 14.6}	&1.16 ($\pm 1.61$)	&3.02 ($\pm 0.34$) \\
    \hline
    \textbf{23.5 / 29.2}	&2.40 ($\pm 0.74$)	&2.55 ($\pm 0.62$) \\
    \hline
    
\end{tabular}
%}
\label{tab:ErrorByFovealSize}
\end{table}
\textbf{Sensitivity to foveal diameter. }We also used the desktop setup to isolate the effect of foveal-region size by varying the foveal diameter setting during VRS texture-map generation and holding all other attack parameters the same as the 120~FPS experiments (including offsets). Participant S0001's gaze data from the ET-DK2 dataset was used five times for each of the three experiment settings. Table~\ref{tab:ErrorByFovealSize} shows that reconstruction accuracy is impacted slightly by fovea size, but the attack remains effective under this realistic range of foveal diameters.

\section{Defense Mechanisms} \label{Defense}
We explored two lightweight but effective machine learning (ML) models to detect the presence of attacks from GPU behavior to provide an initial set of defenses for this novel side-channel vulnerability: a supervised logistic model and an unsupervised K-means clustering model. Both models use time-window summary statistics extracted from GPU metrics (frame rate for desktop and GPU frame time for VR headset) as features. Details of training and evaluation of approaches are provided in Appendix~\ref{AttackDetection}. These models can be used to detect GPU behavior consistent with an ongoing attack to warn users, log events, or trigger countermeasures. While we do not implement countermeasures in this work, we discuss potential approaches later in this section.

\textbf{Logistic regression.} This supervised model represents deployments where labeled attack windows are available. We trained on desktop data with 200~ms positive windows (with HCO scans) and 200~ms null windows (no HCO motion). Although we computed a broad feature set, two features were sufficient: (i) within-window SD and (ii) the outlier proportion (frames outside $\pm 10\%$ of the window mean). A regularized logistic model trained on labeled desktop data achieved an average F1 of $\approx 0.93$ across test scenarios. Because the VR traces use a different metric (GPU frame time) and contain relatively few clear null windows, we treat this model as a desktop baseline and do not retrain it on VR. Instead, we evaluate cross-platform generalization using the K-means approach detailed below.

\textbf{K-means.} The K-means clustering model simulates a deployment scenario with no labeled attack data. We aimed to determine whether clustering could distinguish windows of performance metrics impacted by the attack from normal rendering behavior. We used the same statistical features as in the logistic model and grouped data into 1~s windows from the desktop experiments. Using $k=2$ clusters and the default Euclidean distance metric, the initial model performed worse than logistic regression (F1 score $\approx 0.80$, recall $\approx 0.67$). Inspecting misclassified windows revealed that our original definition of the ``outlier proportion'' feature was overly permissive. Narrowing the outlier threshold from $\pm 10\%$ to $\pm 2\%$ significantly improved performance: the model achieved an F1 score of $\approx 0.99$ and recall of $\approx 1.00$ on the desktop test set. Using this configuration and the same feature definitions, the model generalized well to the VR data, achieving F1 score $= 0.99$ on 1~s windows despite the switch to frame time as the input metric. These results suggest that simple clustering can detect our scan-based attack without labeled data and, when tuned appropriately, can even outperform supervised baselines.

\begin{figure}[!t] %htbp is prefered order of placement of the image: here, top, bottom, seperate page
\centering
\includegraphics[width=0.8\linewidth]{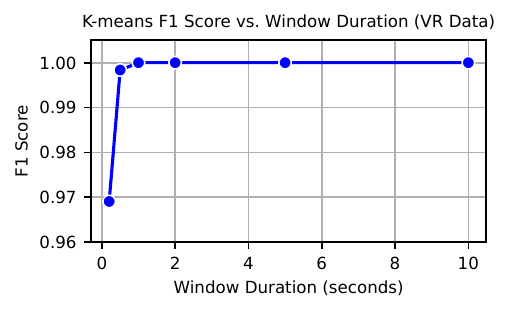}
\caption{VR K-means detection improves as the window duration approaches or exceeds the scan time $T_{\text{scan}} = 500$~ms; performance saturates by $\approx 1$~s.}
\label{fig:KMeansF1VR}
\end{figure}

Fig.~\ref{fig:KMeansF1VR} shows VR attack detection F1 scores as a function of input time window. Performance increases rapidly as the window approaches the scan time $T_{\text{scan}} = 500$~ms and saturates for windows of about 1~s or longer. These results suggest that near-real-time detection is both feasible and highly accurate for our current attack implementation.

\textbf{Countermeasures.} We identified a couple of potential countermeasures that could be used to degrade foveated side-channel attack feasibility. Vertical synchronization, commonly known as VSync, is an optional feature on GPUs that caps the frame rate to match the display's refresh rate to prevent visual artifacts such as screen tearing. Enabling this feature would directly mitigate the use of frame rate as the side-channel metric and cause other side-channels to appear more uniform. However, metrics like GPU render time would still capture relevant information.
 
Another defense strategy is to artificially modulate GPU performance to obscure or mimic attack periodic signature caused by HCO sweeps. For instance, controlled GPU power or load could introduce periodic load fluctuations that align with the attacker's scan period, but time-shifted, making true gaze-induced dips more challenging to discern. This idea is analogous to how deceptive radar jamming can create false aircraft positions by copying a radar signal and emitting it at controlled times~\cite{ali2022electronic}. This defense is well-suited to our periodic sweep: introducing a mismatched periodic load confounds the extremum timing that underpins inference.

Our implementation’s periodic sweep produces a distinctive signature that simple detectors can exploit. Adaptive attackers could vary cadence, randomize scan paths, blend multiple metrics, or interleave scans with content motion to evade fixed detectors. Thus, future work is needed to further understand the breadth of side-channel attacks on foveated rendering as well as the ability to detect and mitigate them.

\section{Related Work } \label{RelatedWork}
We situate this work between VR gaze privacy and GPU side-channel attacks. Prior work shows that eye-tracking data can reveal sensitive attributes and has proposed defenses that regulate direct access to eye-tracking APIs. Separately, GPU side-channel studies use app-accessible performance metrics to infer information about users or rendered content. We bridge these threads: DFR makes rendering load gaze-contingent, and we show that ordinary performance metrics leak gaze even when eye-tracking APIs are unavailable. To our knowledge, no prior work has demonstrated gaze inference on head-mounted displays by exploiting DFR-induced load via app-accessible metrics.

\subsection{Privacy in VR}
Koch highlights unique privacy risks posed by eye-tracking data, including the potential for advertisers and malicious entities to monetize or abuse gaze-derived profiles~\cite{koch2023you}, and emphasizes the current lack of regulation around collection and use of such data. Kröger \textit{et al.} detail sensitive attributes that can be inferred from gaze, such as biometric identifiers, preferences, psychological characteristics, and indicators of mental illness~\cite{kroger2020does}, and argue for stronger legal safeguards for eye-tracking data.

David-John \textit{et al.} describe robust privacy mechanisms to protect eye-tracking data from apps via a gatekeeper API that serves to pass relevant data and features to apps without providing raw gaze data~\cite{david2021privacy}. They also propose a standalone privacy mechanism for features such as foveated rendering that require raw data by adding sample-level noise. However, even if noise is introduced into the foveated rendering pipeline, there is a good chance that fixations can still be inferred and hence some sensitive information can be deduced about the user. 

Steil \textit{et al.} presented a robust way of offering differential privacy that can protect gaze features against re-identification and gender inferences while also providing high performance for eye-tracking applications~\cite{steil2019privacy}. However, their privacy mechanism is applied to aggregated eye-gaze data downstream and would not apply to sample-level DFR or side-channel attacks on GPU metrics. Similarly, the Kal$\epsilon$ido approach provides a spatial differential privacy guarantee on released gaze coordinates that is capable of mitigating privacy attacks but does not protect against GPU-based side-channels.

\subsection{GPU Side-Channel Attacks}
Prior research has shown that GPU side-channel attacks can utilize performance metrics to infer a wide range of sensitive information. Naghibijouybari \textit{et al.} demonstrated that monitoring GPU metrics can reveal substantial information about user activity, including the websites a user visits, activity within those websites, keystroke timing in password fields, and even aspects of neural-network structure~\cite{naghibijouybari2018rendered}. Mai \textit{et al.} demonstrated that it was even possible to geolocate a user based on GPU timing when using WebGL to access certain map websites~\cite{mai2023novel}. 

Zhang \textit{et al.} extended some of the previous work to VR/AR headsets, using CPU and GPU side channels~\cite{zhang2023s}. They monitored both CPU and GPU metrics via API calls to infer information such as running apps, hand gestures used, voice commands that were spoken, virtual keyboard buttons selected, and to a limited extent, how far away bystanders were walking in front of an AR headset.

While prior work has used GPU metrics to infer user actions in both desktop and headset environments, no existing research has explored inferring eye-tracking data from DFR. Our work fills this gap by demonstrating that gaze position can be inferred from DFR-induced workload variations visible in app-accessible performance metrics.
\section{Discussion} \label{Discussion}

\subsection{Preliminary Findings}
\label{Discussion Preliminary Findings}

\textbf{Findings on MQP.}
A key finding is that an unprivileged application on a consumer eye-tracking headset can already reconstruct coarse but non-trivial gaze from DFR and app-accessible performance metrics. In our free-look user study, the scan-based attack produces one $(x,y)$ estimate every 1~s (two 500~ms axis scans) with mean absolute errors of $\approx 4.4^\circ$ (X) and $\approx 2.9^\circ$ (Y) across participants (Table~\ref{tab:leave_one_out_MQP}). Participants did not report noticeable artifacts or quality degradation during the attack, suggesting that the attack can run largely unnoticed under realistic head and eye movements. These results show that on current hardware, DFR-induced load already leaks gaze at a resolution sufficient to localize sustained foveal attention within a few degrees~\cite{mahanama2022eye}.

\textbf{Role of desktop explorations.}
The desktop implementation serves as a controlled testbed for studying how attack accuracy depends on rendering parameters and metric behavior under higher and more stable frame rates. Using real traces from DGaze and ET-DK2, we tuned attack parameters in Godot (including $T_{\text{scan}}$) via calibration trials and then fixed $T_{\text{scan}} = 200$~ms for all reported desktop experiments (Sec.~\ref{sec:desktop_results}). We systematically varied the baseline frame rate between 120, 160, and 200~FPS. For this fixed scan time, increasing the frame rate from 120~FPS to 200~FPS reduced mean error from 3.50$^\circ$/1.59$^\circ$ (X/Y) to 1.19$^\circ$/0.91$^\circ$ (Table~\ref{tab:ErrorByFPSSummary}), and even at 120~FPS the attack already achieved errors comparable to head-mounted eye-tracking noise. These experiments illustrate how frame rate shapes the accuracy and temporal resolution of the attack in a setting where we can tune parameters more freely than on a standalone headset.

\textbf{Temporal-spatial trade-offs and future hardware.}
The desktop experiments highlight a simple but important trade-off. Frame rate determines how finely we can step the HCO across the FOV (HCO position is interpolated each frame), while the scan time $T_{\text{scan}}$ determines how often we can infer a gaze estimate. For a fixed frame rate, shorter scan times improve temporal resolution but reduce the number of samples per scan, which degrades spatial accuracy; for a fixed $T_{\text{scan}}$, higher frame rates improve both spatial sampling and accuracy. On MQP, a 500~ms scan strikes a balance between these factors under current performance constraints. On desktop, where higher frame rates are attainable, preliminary exploration near 500~FPS frame rates showed that a 50~ms scan time was able to yield error magnitudes similar to our 200~ms desktop experiments conducted at 160~FPS. This indicates that given sufficiently high effective frame rates, the same attack could employ shorter scan times resulting in more detailed gaze features. We did not retain the exploratory results at these high frame rates as they are not feasible on current VR headsets, but they are suggestive for future hardware.

Emerging multi-frame-generation techniques, such as NVIDIA’s Blackwell architecture with DLSS4 Multi Frame Generation, are designed to increase effective frame rates by several-fold over traditional rendering~\cite{lin2025dlss}. If similar methods are adopted in VR systems and per-frame performance metrics remain app-accessible, attackers could collect more samples within each scan window and reduce $T_{\text{scan}}$ without sacrificing spatial accuracy. Thus, the temporal resolution we demonstrate here likely represents a conservative lower bound on what DFR-based side channels could achieve on next-generation hardware.

\subsection{Attack Generalizability}
Starting from the MQP study, we next ask how far the attack extends beyond a single headset and scene. Our results indicate that the same basic mechanism generalizes across platforms, game engines, and gaze behaviors.
 
Across platforms, we ran an almost identical scan-based inference pipeline in three settings: MQP (Unity) using \texttt{Time.deltaTime} frame time, a desktop environment (Godot) using frame rate, and a Varjo XR-4 headset (Unreal) using GPU render time. The core pipeline (including scan trajectory, per-scan segmentation, extremum localization, and mapping extremum timing to HCO position) remained unchanged across all three settings. Platform-specific calibration was limited to metric choice and lightweight tuning of HCO density and offset correction. Across these representative systems, we observed comparable accuracy ranges despite differences in hardware configuration, game engine, scene, and DFR pipeline: mean absolute errors remained in the 1--4$^\circ$ range (Sec.~\ref{Results}, Table~\ref{tab:leave_one_out_MQP}, and Table~\ref{tab:ErrorByFPSSummary}). The Varjo XR-4 results, while more limited in sample size, further reinforce that similar leakage arises under a different engine and rendering pipeline observed through a different metric. The clearest trend in our data was with effective frame rate: the two VR headsets produced similar errors operating at similar frame rates (near 70~FPS), and desktop accuracy trended towards the headset accuracy as the frame rate was reduced towards 70~FPS. Taken together, these results suggest that, for these representative systems, no substantial per-device or per-engine tuning is required beyond lightweight calibration.

Additional sensitivity experiments on MQP show that accuracy degrades predictably as the foveation strength decreases. The additional Godot sensitivity experiments show that either doubling or halving the default fovea size resulted in minor accuracy degradation. Notably, these experiments were conducted using the same inference pipeline and previously established calibration settings for each configuration. Therefore, the observed degradation can be attributed to reduced DFR signal strength under modified configurations.

The attack also tolerates realistic gaze behavior. On MQP and Varjo we evaluate natural free-look exploration with head motion. On desktop we replay real traces from DGaze and ET-DK2, which contain natural sequences of fixations and saccades under different tasks and visual content, and we observe similar error ranges across datasets. In all of these conditions, the same HCO design and processing pipeline produces usable gaze estimates without per-user calibration beyond the fixed offsets learned in advance.

Finally, the attack relies only on app-accessible performance metrics that are already exposed~\cite{zhang2023s}. As long as DFR is active and these metrics remain available, our results suggest that scan-based gaze inference remains a plausible threat beyond the specific hardware and scenes we study.

\subsection{Privacy Implications of Coarse Gaze}
On MQP, our reconstruction yields a relatively low-rate gaze series: with $T_{\text{scan}} = 500$~ms per axis, we obtain one $(x,y)$ estimate roughly every second. On desktop, shorter scans and higher frame rates yield estimates every 400~ms. In both cases, each estimate is derived from a scan window comparable to a typical fixation duration (200--300~ms)~\cite{lappi2016eye}. Prior work links fixation location and duration to the deployment of visual attention and deeper cognitive processing~\cite{mahanama2022eye}. As a result, even coarse gaze recovery allows an attacker to identify which regions or objects in a scene received sustained foveal attention over time.

For many realistic attack goals, this level of information is already privacy-sensitive. For example, in a product gallery or news feed, sampling gaze every 0.5--1~s is sufficient to determine which item a user dwelled on before acting and for roughly how long, supporting detailed profiling of interests or systematic experimentation with alternative content layouts. In social or collaborative VR, coarse gaze can reveal which avatars or bystanders a user repeatedly looks at, or which panels of a multi-window workspace attract sustained or frequent attention, even if rapid saccades between them are not resolved.

These findings complement prior work on gaze privacy that assumes high-rate eye tracking. Even without access to eye-tracking APIs, and even at low temporal resolution, DFR-based side channels expose enough information about user focus to undermine privacy mechanisms that focus on restricting access to or sanitizing explicit gaze streams.

\subsection{Limitations}

\textbf{HCO design and inference model.}
Our cylinder-based HCO yields good accuracy but has limits. As a thin on-screen band, its fovea overlap is brief at lower frame rates, reducing samples per pass. We employed a simple linear correction model in the VR attack setting, and more complex transformations are likely to improve accuracy.

\textbf{Temporal resolution.}
Our scanning-based HCO design presents a trade-off between scan time and gaze inference accuracy. Alternative HCO layouts or scan methods may need to be employed to reduce this tradeoff on current headsets. We also expect that learning-based methods could upsample our coarse gaze series to a higher temporal resolution in future work~\cite{jiao2023supreyes}.

\textbf{Link to downstream privacy.}
Our scope was to demonstrate accurate gaze reconstruction and release a small dataset. A larger VR dataset and an evaluation that connects reconstructed gaze trajectories to concrete downstream inferences are needed to quantify end-to-end privacy risk and to validate the coarse-gaze attacks we discuss above in realistic application scenarios.
\section{Conclusion} \label{Conclusion}
In this paper, we demonstrated a novel side-channel attack that bypasses eye-gaze privacy mechanisms proposed in other works by exploiting foveated rendering and commonly accessible GPU performance metrics. We showed that such an attack is feasible across various platforms, rendering pipelines, and game engines, and can accurately infer gaze coordinates based solely on a single performance metric provided it is correlated with GPU workload. We also provided an initial mechanism for detecting these types of side-channel attacks, but mitigations and consideration of more complex attacks are critical directions for future work. Our research artifacts are available at:
\url{https://version.cs.vt.edu/privateeye/eyespy}

\section*{Ethics Considerations}
We conducted a stakeholder-focused ethics analysis guided by the Menlo Report, considering study participants, VR end users affected by gaze inference, bystanders in AR/VR contexts, platform and engine vendors, and the research community. We did not target or exclude populations based on protected characteristics; instead, we evaluate a system-level side channel that arises across widely deployed headsets and engines, and design mitigations intended to protect the broad user base. Our VR user studies were approved by the IRB, posed negligible risk, and obtained informed consent from all participants, who reported no adverse symptoms from their brief participation.

Given the nature of our attack work, we notified both Varjo and Meta's Bug Bounty program for vulnerability disclosure prior to paper submission. We held follow-up conversations with Varjo who determined that the exploit, while relevant, did not fit their security threat model. Varjo does not target a commercial user base, instead providing a research-grade system to customers who use the headset for enterprise applications or research studies. Thus, these users or research participants would be aware that eye-tracking data is being recorded through informed consent and is available through the API. Second, Varjo best practices are to maintain an isolated desktop for running validated software on the headset which would prevent malicious applications from embedding our attack within third-party software. We did not hear back from Meta regarding our ethical vulnerability disclosure.

\section*{LLM Usage Considerations}

LLMs were used for editorial purposes in this manuscript, and all outputs were inspected by the authors to ensure accuracy and originality.

\section*{Acknowledgments}
The authors acknowledge funding from the National Science Foundation (CNS-2350116) and the Commonwealth Cyber Initiative.

\bibliographystyle{IEEEtran}
\bibliography{Sections/References}

\appendices

\section{Desktop Calibration and Additional Results}
\label{app:desktop}
\subsection{Desktop Scan Time and Spatial Offsets}
\label{app:scan-offset}
We calibrated the desktop pipeline in the Sponza scene under a Center-gaze condition without backdrops so that baseline frame rate depended only on the HCO cylinder count. We swept scan times (100--1000~ms) while varying cylinder count to cover a wide frame rate range, then computed mean and SD of the per-scan angular error for each setting. We chose a 200~ms scan time because it was the fastest setting that yielded low and stable SD across runs when FPS was \(\ge\)120~FPS. Although longer scans reduced mean error, they did not improve SD; since a constant spatial bias can be compensated for with an offset and variability cannot, we prioritized lower SD.

To obtain offsets, we took the mean per-axis error from the 200~ms Center-gaze runs at each frame rate (120/160/200~FPS). These values were recorded in Table~\ref{tab:FPS_Offsets} and subtracted as offsets from each coordinate estimate in subsequent desktop experiments. The errors for these experiments can be found in Tables~\ref{tab:ErrorWithOffset120FPS},~\ref{tab:ErrorWithOffset160FPS}, and~\ref{tab:ErrorWithOffset200FPS}.
\begin{table}[!t]
\centering
\caption{Average offsets for 200~ms scan calculated from the Center calibration experiments.}
%\resizebox{\columnwidth}{!}{
\begin{tabular}{|c|c|c|c|c|}
\hline
\textbf{FPS}    & \textbf{Mean Offset X ($^\circ$)} & \textbf{Mean Offset Y ($^\circ$)}  \\ \hline
120             & 8.64$^\circ$                      & 4.80$^\circ$                       \\ \hline
160             & 6.79$^\circ$                      & 3.98$^\circ$                       \\ \hline
200             & 5.54$^\circ$                      & 3.28$^\circ$                       \\ \hline

\end{tabular}
%}
\label{tab:FPS_Offsets}
\end{table}
\begin{table}[!t]
\centering
\caption{Mean and SD of Errors for Different Eye-Gaze Conditions (120~FPS)}
\resizebox{\columnwidth}{!}{%
\begin{tabular}{|r|c|c|c|c|}
    \hline
    \textbf{Eye-Gaze Conditions} & \textbf{Avg Error X$^\circ$} & \textbf{SD X$^\circ$} & \textbf{Avg Error Y$^\circ$} & \textbf{SD Y$^\circ$} \\
    \hline
    Center (Calibration) & 2.37 & 3.32 & 1.32 & 1.63 \\
    \hline
    % Grid & 14.44 & 23.97 & 1.67 & 2.43 \\
    % \hline
    DGaze & 2.27 & 3.0 & 1.39 & 1.69 \\
    \hline
    ET-DK2 S0001 & 3.17 & 5.72 & 1.79 & 2.81 \\
    \hline
    ET-DK2 S0002 & 2.26 & 3.39 & 1.34 & 1.61 \\
    \hline
    ET-DK2 S0003 & 2.61 & 3.86 & 1.46 & 1.78 \\
    \hline
    ET-DK2 S0004 & 2.79 & 5.14 & 1.55 & 1.92 \\
    \hline
    ET-DK2 S0005 & 2.7 & 4.99 & 2.12 & 3.51 \\
    \hline
    ET-DK2 S0006 & 2.23 & 2.9 & 1.38 & 1.69 \\
    \hline
    ET-DK2 S0008 & 3.77 & 9.47 & 1.74 & 2.83 \\
    \hline
    ET-DK2 S0009 & 3.46 & 8.36 & 1.47 & 2.3 \\
    \hline
    ET-DK2 S0010 & 2.47 & 3.08 & 1.6 & 2.38 \\
    \hline
    ET-DK2 S0011 & 3.16 & 6.84 & 1.4 & 1.7 \\
    \hline
    ET-DK2 S0012 & 2.59 & 5.63 & 1.35 & 1.62 \\
    \hline
    ET-DK2 S0013 & 5.47 & 13.73 & 1.78 & 2.74 \\
    \hline
    ET-DK2 S0014 & 2.1 & 3.12 & 1.42 & 1.92 \\
    \hline
    ET-DK2 S0015 & 3.09 & 5.31 & 2.34 & 4.43 \\
    \hline
    ET-DK2 S0016 & 2.45 & 4.16 & 1.43 & 1.78 \\
    \hline
    ET-DK2 S0017 & 4.51 & 11.21 & 1.94 & 3.25 \\
    \hline
    ET-DK2 S0018 & 2.99 & 6.22 & 1.43 & 1.74 \\
    \hline
    ET-DK2 S0019 & 2.52 & 3.35 & 1.39 & 1.78 \\
    \hline
    \textbf{ET-DK2 Averages:}	&\textbf{3.02}	&\textbf{5.92}	&\textbf{1.61}	&\textbf{2.32} \\
    \hline
\end{tabular}
}
\label{tab:ErrorWithOffset120FPS}
\end{table}
\begin{table}[!t]
\centering
\caption{Mean and SD of Errors for Different Eye-Gaze Conditions (160~FPS)}
\resizebox{\columnwidth}{!}{
\begin{tabular}{|r|c|c|c|c|}
    \hline
    \textbf{Eye-Gaze Conditions} & \textbf{Avg Error X$^\circ$} & \textbf{SD X$^\circ$} & \textbf{Avg Error Y$^\circ$} & \textbf{SD Y$^\circ$} \\
    \hline
    Center (Calibration) & 0.92 & 1.18 & 0.53 & 0.79 \\
    \hline
    % Grid & 2.06 & 7.11 & 0.87 & 1.12 \\
    % \hline
    DGaze & 1.45 & 2.08 & 0.83 & 1.01 \\
    \hline
    ET-DK2 S0001 & 2.66 & 5.44 & 1.67 & 3.08 \\
    \hline
    ET-DK2 S0002 & 1.45 & 1.68 & 1.07 & 1.84 \\
    \hline
    ET-DK2 S0003 & 1.46 & 1.73 & 0.89 & 1.18 \\
    \hline
    ET-DK2 S0004 & 1.73 & 3.2 & 1.15 & 1.71 \\
    \hline
    ET-DK2 S0005 & 1.82 & 3.46 & 1.51 & 2.48 \\
    \hline
    ET-DK2 S0006 & 1.35 & 1.51 & 0.99 & 1.41 \\
    \hline
    ET-DK2 S0008 & 1.4 & 1.73 & 1.35 & 2.02 \\
    \hline
    ET-DK2 S0009 & 1.46 & 1.8 & 1.0 & 1.34 \\
    \hline
    ET-DK2 S0010 & 1.55 & 1.88 & 1.19 & 1.74 \\
    \hline
    ET-DK2 S0011 & 1.48 & 1.83 & 0.93 & 1.29 \\
    \hline
    ET-DK2 S0012 & 1.31 & 1.56 & 0.91 & 1.35 \\
    \hline
    ET-DK2 S0013 & 2.82 & 8.41 & 1.49 & 2.27 \\
    \hline
    ET-DK2 S0014 & 1.63 & 3.55 & 0.98 & 1.19 \\
    \hline
    ET-DK2 S0015 & 2.14 & 4.47 & 1.61 & 2.98 \\
    \hline
    ET-DK2 S0016 & 1.3 & 1.77 & 0.9 & 1.23 \\
    \hline
    ET-DK2 S0017 & 1.75 & 2.39 & 1.35 & 3.03 \\
    \hline
    ET-DK2 S0018 & 1.64 & 2.6 & 1.02 & 1.53 \\
    \hline
    ET-DK2 S0019 & 1.32 & 1.54 & 1.18 & 2.21 \\
    \hline
    \textbf{ET-DK2 Averages:}	&\textbf{1.68}	&\textbf{2.81}	&\textbf{1.18}	&\textbf{1.88} \\
    \hline
\end{tabular}
}
\label{tab:ErrorWithOffset160FPS}
\end{table}

\begin{table}[!t]
\centering
\caption{Mean and SD of Errors for Different Eye-Gaze Conditions (200~FPS)}
\resizebox{\columnwidth}{!}{%
\begin{tabular}{|r|c|c|c|c|}
    \hline
    \textbf{Eye-Gaze Conditions} & \textbf{Avg Error X$^\circ$} & \textbf{SD X$^\circ$} & \textbf{Avg Error Y$^\circ$} & \textbf{SD Y$^\circ$} \\
    \hline
    Center (Calibration) & 0.77 & 0.65 & 0.4 & 0.76 \\
    \hline
    % Grid & 0.98 & 2.23 & 0.69 & 1.7 \\
    % \hline
    DGaze & 1.1 & 1.09 & 0.73 & 0.9 \\
    \hline
    ET-DK2 S0001 & 1.94 & 4.62 & 1.46 & 3.03 \\
    \hline
    ET-DK2 S0002 & 1.13 & 1.33 & 0.76 & 1.05 \\
    \hline
    ET-DK2 S0003 & 1.09 & 1.16 & 0.7 & 0.97 \\
    \hline
    ET-DK2 S0004 & 1.07 & 1.36 & 0.9 & 1.39 \\
    \hline
    ET-DK2 S0005 & 1.25 & 1.61 & 1.11 & 1.66 \\
    \hline
    ET-DK2 S0006 & 1.1 & 1.15 & 0.89 & 1.3 \\
    \hline
    ET-DK2 S0008 & 1.07 & 1.21 & 1.05 & 1.63 \\
    \hline
    ET-DK2 S0009 & 1.17 & 1.31 & 0.83 & 1.06 \\
    \hline
    ET-DK2 S0010 & 1.31 & 2.23 & 0.98 & 1.54 \\
    \hline
    ET-DK2 S0011 & 1.09 & 1.13 & 0.85 & 1.14 \\
    \hline
    ET-DK2 S0012 & 1.08 & 1.24 & 0.72 & 1.08 \\
    \hline
    ET-DK2 S0013 & 1.38 & 3.38 & 1.08 & 1.77 \\
    \hline
    ET-DK2 S0014 & 1.09 & 1.18 & 0.79 & 1.07 \\
    \hline
    ET-DK2 S0015 & 1.47 & 2.29 & 1.51 & 3.33 \\
    \hline
    ET-DK2 S0016 & 1.09 & 1.23 & 0.84 & 1.16 \\
    \hline
    ET-DK2 S0017 & 1.54 & 2.0 & 1.08 & 1.83 \\
    \hline
    ET-DK2 S0018 & 1.22 & 1.42 & 0.8 & 1.03 \\
    \hline
    ET-DK2 S0019 & 1.04 & 1.06 & 0.95 & 1.56 \\
    \hline
    \textbf{ET-DK2 Averages:}	&\textbf{1.23}	&\textbf{1.72}	&\textbf{0.96}	&\textbf{1.53} \\
    \hline
\end{tabular}
}
\label{tab:ErrorWithOffset200FPS}
\end{table}

\subsection{Attack Detection}
\label{AttackDetection}
We evaluate two lightweight detectors: (i) a \emph{supervised} logistic regression trained on labeled windows of scan and no-scan periods, and (ii) an \emph{unsupervised} K-means model operating on fixed-duration windows with no labels. Each window is summarized by simple statistics of the rendering metric: standard deviation (SD), skewness, kurtosis, range, interquartile range (IQR), and the outlier proportion (samples outside $\pm 2\%$ of the window mean) after SG smoothing (window = 15, poly = 5). For the logistic model we stratify a 70/30 train/test split across 120/160/200~FPS frame rates and conditions (Attack/NoAttack balanced). A two-feature variant using only \{SD, outlier proportion\} performs nearly as well as when using the full set of features; on desktop logistic regression attains F1 $\approx 0.93$ across conditions.

The label-free K-means detector targets the deployment case where neither the scan period nor training data are known. Using the same features and smoothing, tightening the outlier threshold from $\pm 10\%$ to $\pm 2\%$ (after smoothing) eliminates most false negatives. Performance saturates once the window length meets or exceeds the scan time: on desktop, K-means achieves F1 $\approx 0.99$ for windows $\ge T_{\text{scan}}$; for both MQP (frame time) and Varjo XR-4 (GPU render time), the model attains an F1 of $0.99$ for $\ge 1$~s windows with the same parameters. Because it requires no labels and can run with \{SD, outlier proportion\} alone, this clustering approach is practical for online screening and complements more heavyweight defenses.

\section{Varjo Implementation and MQP Supplemental}
\label{app:varjo}
\label{app:VR}

\begin{figure}[!t]
\centering
\subfloat[]{\includegraphics[width=0.47\linewidth]{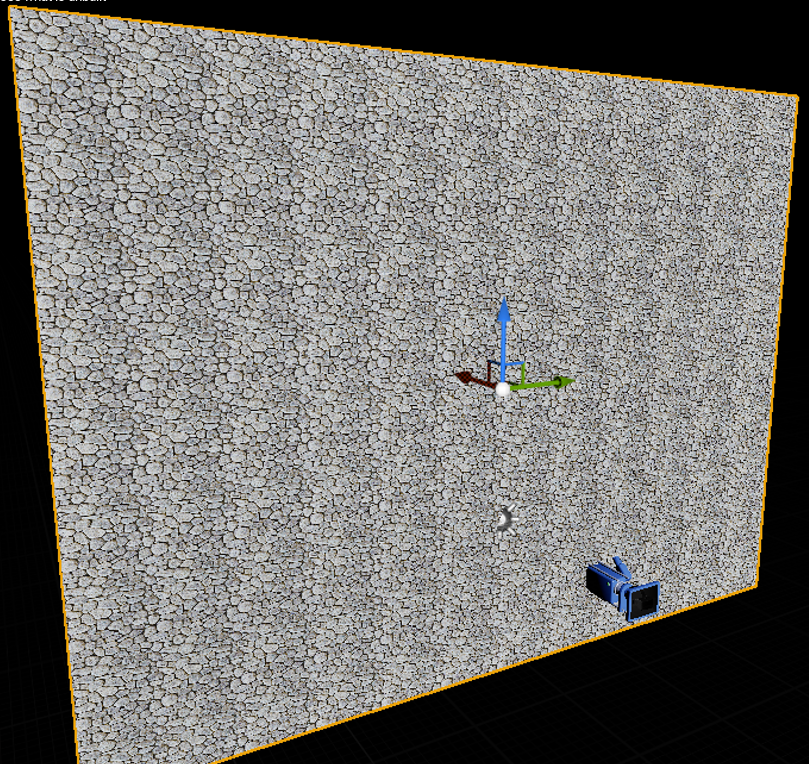}\label{fig:subfigRW}}
\hfill
\subfloat[]{\includegraphics[width=0.45\linewidth]{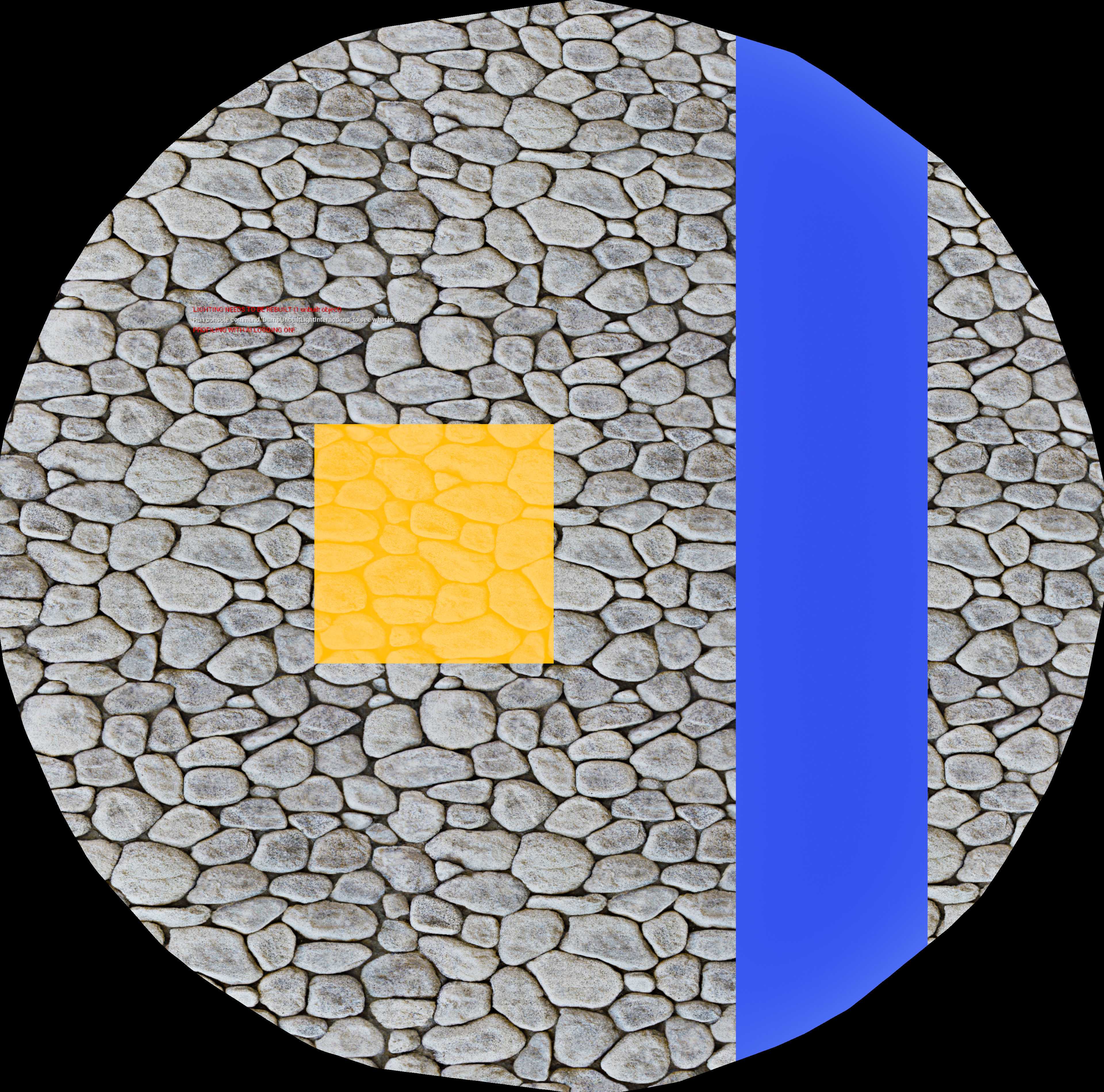}\label{fig:subfigVR}}
\caption{(a) Unreal editor scene. (b) Left-eye view showing the foveal region. The attack HCO and square foveal viewport are invisible to the user and are shown here in blue and yellow respectively for illustration.}
\label{fig:unreal_vr_panels}
\end{figure}
We implemented our Varjo attack within the Unreal game engine, as it is an industry standard for VR experiences and has good compatibility with VR displays for logging ground-truth gaze data. We chose Varjo XR-4 primarily for its integrated 200~Hz eye tracker, enabling dense ground-truth gaze data.

We use OpenXR quad-view foveation~\cite{openxr_quadviews} rather than VRS, as Unreal’s available VRS path at the time was fixed-foveation and not stereo-aware. Like the MQP, we employ a single perspective camera and move the HCOs close to the camera’s near plane to help mitigate distortion effects. 

\subsection{VR Environment}
Due to computer availability when running the VR study, a different but similar computer was utilized. We performed the experiments on an Alienware Aurora R16 desktop computer with an Intel Core i9-14900F, 64~GB RAM, and an NVIDIA RTX 4090 GPU. The Unreal Engine (5.4.2) on Windows 11 was used to render a simple Rock Wall attack scene to the Varjo XR-4 headset. The Rock Wall scene consists of a textured ground plane and a single light source. We used this low-complexity environment to isolate the effects of our attack when using real-time eye-tracking data from the headset instead of pre-recorded traces and due to a lack of open-source graphics benchmark scenes configured in Unreal for the Varjo headset.

Fig.~\ref{fig:subfigRW} shows the Rock Wall VR attack scene while Fig.~\ref{fig:subfigVR} shows the view of the HCO and foveal region for one eye. The Varjo XR-4 provides a FOV of approximately 120$^\circ$ in the X-dimension and 105$^\circ$ in the Y-dimension.

\subsection{Varjo XR-4 Foveated Rendering}
We first explored NVIDIA VRS-based foveation in Unreal, but available Unreal Engine 5 paths were fixed-foveation and not stereo-aware. We therefore used Varjo’s quad-view foveated rendering~\cite{openxr_quadviews}, which keeps the shading rate at 1$\times$1 but renders a high-resolution foveal viewport and a lower-resolution peripheral viewport for each eye. The foveal viewport covers a square region extending 15$^\circ$ around the current gaze point (Fig.~\ref{fig:subfigVR}). We set Varjo image quality to ``High'', where the foveal view is 1472$\times$1472 and the peripheral view is 2100$\times$2084~\cite{varjo_quality_settings}.

\subsection{Varjo XR-4 Attack}
The Varjo XR-4 attack was implemented the same as the MQP, just with minor differences in HCO size and quantity. The Unreal editor provides direct access to performance metrics during rendering through integrated debug tools. The typical frame rate metric was not ideal since the CPU process added delay to the frame rate. We instead used Unreal's GPU render time metric since it more closely reflected GPU workload. We inverted the GPU Time ($\frac{1}{\text{GPU render time}}$) to get a more natural frame rate type of metric. 

\subsection{Varjo XR-4 Implementation}
\label{sec:varjo_user_study}
\begin{figure}[!t]
\centering
\subfloat[]{%
  \includegraphics[width=0.49\linewidth]{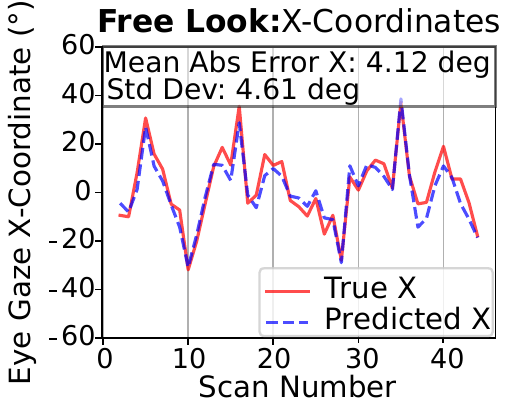}%
  \label{fig:subfig2}}
\hfill
\subfloat[]{%
  \includegraphics[width=0.49\linewidth]{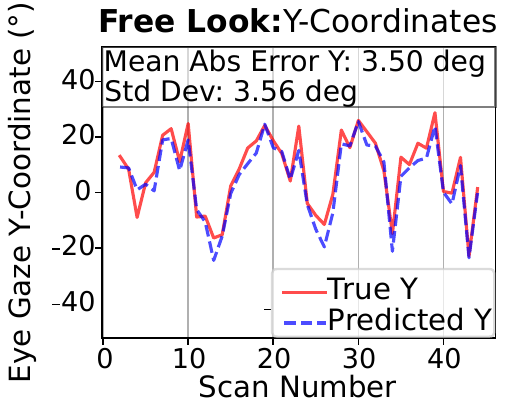}%
  \label{fig:subfig4}}
\caption{Predicted and ground-truth VR gaze positions for Participant~1 when attack parameters are optimized on the remaining participants. The predicted gaze closely follows the ground-truth gaze across all conditions. 
}
\label{fig:vrPrelimResultsFree}
\end{figure}
We conducted an IRB-approved user study with six participants to evaluate how well our attack performs under real-time eye-tracking conditions on the Varjo XR-4 headset. All participants provided informed consent. During the session, each participant completed a 45~s Free Look task where the participant naturally explored the scene without prompts, allowing for realistic viewing. Participants completed the Simulator Sickness Questionnaire (SSQ)~\cite{kennedy1993simulator} both before and after the session. None of our participants reported experiencing symptoms during their sessions. 

We apply the preprocessing and gaze-estimation pipeline described in Sec.~\ref{Implementation} and Algorithm~1, using $\frac{1}{\text{GPU render time}}$ as the performance metric. Each scan window is processed independently to produce an initial gaze estimate ($P_i$) along the scanned axis, and we set $T_{\text{scan}} = 500$~ms per axis to accommodate the effective frame rates on Varjo XR-4.

To compensate for increased errors at higher eccentricities, we introduced a non-linear correction function that scaled with the eccentricity of the initial prediction. This adjustment improved accuracy across the full visual field. We optimized the following equation to minimize mean absolute error when computing attack coefficients $ P_f = P_i - C + A \cdot \tan(K(P_i - C)),$ where $P_f$ is the final predicted gaze position, $C$ is a scalar offset, and $A \cdot \tan(K(P_i - C))$ represents a scaling term based on how far the offset gaze estimate is from the center of the FOV.

Our attack used data from the VR user study to optimize coefficients and evaluate attack performance. The three coefficients $C$, $A$, and $K$ were optimized independently for X- and Y-scans for a given set of training data. The coefficients were optimized in a modified grid search by first setting $A$ and $K$ to zero and identifying the optimal value of $C \in [15^\circ, 30^\circ]$ with a step size of $0.1^\circ$. The value of $C$ was then fixed to the value that produced the lowest mean absolute gaze error. Next, pairs of $A,K$ values were sampled to find optimal values within the ranges $A \in [10, 40]$ and $K \in [1, 4]$ with a step size of $0.2$. 

Table~\ref{tab:leave_one_out} presents the LOOCV accuracy for each axis across the six participants. Coefficients are optimized on five participants and evaluated on the held-out participant, then averaged across folds.
\begin{table}[!t]
\centering
\caption{Mean and SDs of absolute gaze error across all six folds in a LOOCV evaluation}
\renewcommand{\arraystretch}{1.1}  % increase row height a bit (default 1.0)
\begin{tabular}{|c|c|c|}
\hline
\textbf{Condition} & \textbf{X Error ($^\circ$)} & \textbf{Y Error ($^\circ$)} \\ \hline
%\textbf{Grid}      & $4.51\,(\pm\,4.60)$ & $3.19\,(\pm\,4.17)$ \\ \hline
\textbf{Free Look} & $3.98\,(\pm\,4.81)$ & $3.24\,(\pm\,4.18)$ \\ \hline
\end{tabular}
\label{tab:leave_one_out}
\end{table}
Errors ranged up to an average of 4.51$^\circ$~(X) and 3.19$^\circ$~(Y). Fig.~\ref{fig:vrPrelimResultsFree} illustrates how the inferred gaze position follows the ground truth data by plotting data from Participant 1 with coefficients trained on all other participants. The magnitude of errors and SDs suggests room for improvement in predicting gaze position; however, we consider these findings to show the attack is feasible. 

\subsection{MQP Participant Data}
Results for the 15 MQP participants are in Table~\ref{tab:LOOCVErrorStats}. 

%\begin{table}[h]
\begin{table}[!t]
\centering
\caption{Mean and SD of Prediction Errors per LOOCV Fold (MQP, 70~FPS)}
\resizebox{\columnwidth}{!}{%
\begin{tabular}{|r|c|c|c|c|}
    \hline
    \textbf{Fold} & \textbf{Avg Error X$^\circ$} & \textbf{Avg SD X$^\circ$} & \textbf{Avg Error Y$^\circ$} & \textbf{Avg SD Y$^\circ$} \\
    \hline
     1  & 3.97 & 5.63 & 5.45 & 6.63 \\ \hline
     2  & 3.99 & 5.79 & 2.44 & 2.05 \\ \hline
     3  & 3.58 & 4.81 & 3.31 & 4.49 \\ \hline
     4  & 5.29 & 6.84 & 2.25 & 2.43 \\ \hline
     5  & 5.08 & 8.36 & 1.88 & 2.20 \\ \hline
     6  & 4.35 & 4.98 & 1.34 & 1.70 \\ \hline
     7  & 3.64 & 5.03 & 2.34 & 3.42 \\ \hline
     8  & 5.44 & 6.41 & 2.86 & 5.10 \\ \hline
     9  & 3.95 & 5.06 & 1.53 & 1.99 \\ \hline
    10  & 3.51 & 4.62 & 3.08 & 3.18 \\ \hline
    11  & 4.10 & 5.47 & 2.29 & 3.07 \\ \hline
    12  & 4.36 & 6.18 & 1.74 & 2.35 \\ \hline
    13  & 4.54 & 6.77 & 5.83 & 8.96 \\ \hline
    14  & 4.79 & 6.57 & 3.30 & 4.36 \\ \hline
    15  & 4.76 & 5.88 & 3.52 & 2.07 \\ \hline
    \textbf{Average} & \textbf{4.36} & \textbf{5.89} & \textbf{2.88} & \textbf{3.60} \\ \hline
\end{tabular}
}
\label{tab:LOOCVErrorStats}
\end{table}

\newpage % The Meta-Review should at least start on a new column

% Use \appendices and not \appendix due to IEEEtran.cls quirks
%\appendices % if not used earlier

\section{Meta-Review}

The following meta-review was prepared by the program committee for the 2026
IEEE Symposium on Security and Privacy (S\&P) as part of the review process as
detailed in the call for papers.

\subsection{Summary}
This paper demonstrates that Dynamic Foveated Rendering (DFR) technology can be used as a side channel to infer user gaze location, even if eye tracking data would not be directly available to the adversary. The VR app developer implements horizontal and vertical scans to the video feed. The scans are not visible to the human user, but when the scan hits the foveated area, which is rendered with higher precision, the computing load increases. By monitoring GPU metrics and correlating it with scan timings, the adversary can infer gaze location. 

\subsection{Scientific Contributions}
\begin{itemize}
\item Independent Confirmation of Important Results with Limited Prior Research.
\item Creates a New Tool to Enable Future Science.
\item Identifies an Impactful Vulnerability.
\item Provides a Valuable Step Forward in an Established Field.
\end{itemize}

\subsection{Reasons for Acceptance}
\begin{enumerate}
\item This paper identifies a novel side channel for VR apps based on DFR. This side channel allows a malicious VR app developer to infer user gaze without direct access to eye tracking data. Thus the paper nicely demonstrates that existing defenses like API gating alone may be insufficient to protect the privacy of VR app users.
\item The paper presents a concrete attack pipeline that infers gaze points by correlating application-visible performance metrics with known object movements. The use of visually unobtrusive, transparent HCOs as probing signals makes the side channel exploitable under realistic UI constraints. The paper also provides strong cross-layer insight: Links human vision, rendering optimization, hardware metrics, and privacy leakage.
\item Evaluation on Meta Quest Pro (Unity), Varjo XR-4 (Unreal), and desktop pipelines (Godot) demonstrates that the phenomenon is not a single-device artifact.
\end{enumerate}

\end{document}